\documentclass[aps,prl,showpacs,twocolumn,superscriptaddress,floatfix,preprintnumbers]{revtex4-2}
\usepackage{graphicx}
\usepackage{dcolumn}
\usepackage{bm}
\usepackage{braket}
\usepackage{hyperref}
\usepackage{amsmath}
\usepackage{amsthm,amssymb}
\usepackage{mathrsfs}
\usepackage{color}
\hypersetup{colorlinks=true, linkcolor=blue, anchorcolor=blue, urlcolor=blue, citecolor=blue}
\renewcommand{\eqref}[1]{\hyperref[#1]{(}\ref{#1}\hyperref[#1]{)}}

\begin{document}

\title{Tunable Magnonic Chern Bands and Chiral Spin Currents in Magnetic Multilayers}

\author{Zhongqiang Hu}
\email{zhongqhu@mit.edu}
\affiliation{Department of Electrical Engineering and Computer Science, Massachusetts Institute of Technology, Cambridge, MA 02139, USA}

\author{Liang Fu}
\affiliation{Department of Physics, Massachusetts Institute of Technology, Cambridge, MA 02139, USA}

\author{Luqiao Liu}
\affiliation{Department of Electrical Engineering and Computer Science, Massachusetts Institute of Technology, Cambridge, MA 02139, USA}

\date{\today}

\begin{abstract}
	
Realization of novel topological phases in magnonic band structures represents a new opportunity for the development of spintronics and magnonics with low power consumption. In this work, we show that in antiparallelly aligned magnetic multilayers, the long-range, chiral dipolar interaction between propagating magnons generates bulk bands with non-zero Chern integers and magnonic surface states carrying chiral spin currents. The surface states are strictly localized and can be easily toggled between non-trivial and trivial phases through an external magnetic field. The realization of chiral surface spin currents in this dipolarly coupled heterostructure represents a magnonic implementation of the coupled wire model that has been extensively explored in electronic systems. Our work presents an easy-to-implement system for realizing topological magnonic surface states and low-dissipation spin current transport in a tunable manner.

\end{abstract}

\maketitle

Exploration of novel topological phases in quantum matter has become one of the central topics in nowadays' condensed matter research, opening up avenues towards electronics with high speed and low power consumption \cite{Hasan2010, Qi2011, Ando2015, Sato2017, Armitage2018}. Beyond electronic systems, recently topological phases have also been generalized to various bosonic systems, including phononic \cite{Wang2015, He2018} and photonic \cite{Wu2015, Lu2016} crystals. Magnon, the quantized collective excitation of localized spins, represents a promising candidate for efficient spin transport \cite{Kajiwara2010, Cornelissen2015, Han2019}. However, the inevitable scattering between magnons greatly limits the coherence length of spin waves, preventing long-range spin signal transfer \cite{Turner1960, Raquet2002, Boona2014}. The formation of topological magnonic surface states that provide spin current channels with suppressed scattering is therefore of great importance for realizing low-dissipation magnonic devices, which has been proposed in several theoretical works \cite{Zhang2013, Mook2014, Owerre2016, Kim2016, Hirosawa2020, Shindou2013, Li2018, Wang2020}. Nevertheless, these proposals require materials with either special crystal symmetries \cite{Zhang2013, Mook2014, Owerre2016, Kim2016, Hirosawa2020} or artificially modulated structures that demand advanced nanofabrication techniques \cite{Shindou2013, Li2018, Wang2020}, both of which bring in difficulties for experimental realization.

In this letter, we theoretically study the magnonic band structure and corresponding topological properties of antiparallelly aligned magnetic multilayers. We find that the long-range dipolar interaction between propagating magnons is chiral in nature, whose strength depends on the wave vector direction and therefore breaks time-reversal symmetry (TRS). This chiral dipolar interaction correlates the sublattice and momentum degrees of freedom, playing a similar role as the spin-orbit coupling does in electronic systems. Consequently, it generates bulk bands with non-zero Chern integers and ultra-localized magnonic surface states that carry chiral spin currents. Through an external magnetic field, the existence of topologically non-trivial surface states can be switched on and off, which therefore provides a tunable and efficient way for transferring spin angular momenta in this synthetic antiferromagnetic heterostructure.

\begin{figure}[t]
	\centering
	\includegraphics[width=1\columnwidth]{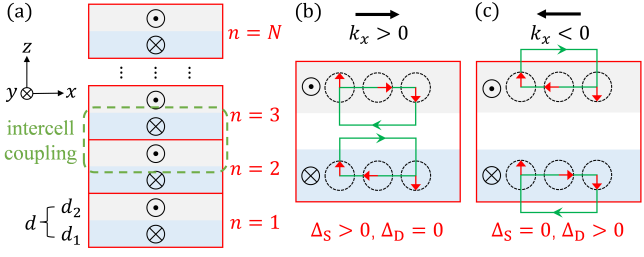}
	\caption{(a) Scheme of antiparallelly aligned magnetic multilayers. The blue (grey) blocks with the thickness of $d_{1(2)}$ form the $1$st ($2$nd) sublattice. The equilibrium moments in the $1$st ($2$nd) sublattice are parallel to the $+y$ ($-y$) direction, represented by $\otimes$ ($\odot$) symbols. The red, solid frames correspond to unit cells (indexed by $n = 1, 2, ..., N$) with the thickness of $d = d_1 + d_2$. The green, dotted frames select two layers belonging to neighboring cells with possible intercell dipolar coupling. (b) Scheme of dipolar fields (green lines and arrows) generated by propagating magnons with $k_x > 0$ in two layers within a cell. The red arrows represent the deviations of magnetic moments from their equilibrium orientations, which propagates from left to right as a function of time (the traces are represented by black, dotted circles). Here, the two layers experience finite dipolar fields from each other, corresponding to $\Delta_\mathrm{S} > 0$. Meanwhile, they experience zero dipolar fields from magnons in the layers of neighboring cells, corresponding to $\Delta_\mathrm{D} = 0$. (c) Scheme of dipolar fields generated by propagating magnons with $k_x < 0$ for two layers within a cell, corresponding to $\Delta_\mathrm{S} = 0$ and $\Delta_\mathrm{D} > 0$.}
	\label{fig:1}
\end{figure}

We focus on the magnonic band structure in magnetic multilayers shown in Fig. \hyperref[fig:1]{1(a)}, where the neighboring layers possess antiparallel equilibrium magnetic moments due to antiferromagnetic interfacial exchange, as demonstrated in several recent experiments \cite{Klingler2018, Chen2018, Fan2020}. An external magnetic field applied along the $y$ axis always aligns the higher- (lower-) moment layers parallel (antiparallel) to it. The layers with the same equilibrium moment orientations have entirely identical properties including their material composition and thickness, providing the system with periodicity along the $z$ axis, i.e., film growth direction, and allowing  us to define unit cells with thickness of $d = d_1 + d_2$, as shown in red frames in Fig. \hyperref[fig:1]{1(a)}.

We start by considering the simplified case, where each layer is a infinitely long strip along the $x$ axis and the magnetic moment distribution along the $y$ and $z$ axes within the same layer is uniform, i.e., the lowest-order standing wave mode. As a result, magnons are confined to transport along the $x$ axis in each individual layer. We denote the annihilation operator for a circularly polarized magnon with wave vector $k_x$ in the $j$th ($j = 1, 2$ for sublattice index) layer of the $n$th ($n = 1, 2, ..., N$ for cell index) cell as $b_{jn,k_x}$. Neglecting the interlayer interactions, we can write the magnonic Hamiltonian for the multilayers in Fig. \hyperref[fig:1]{1(a)} as $\hat{\mathcal{H}}_0 = \sum_{k_x, j, n} \omega_j b_{jn,k_x}^\dagger b_{jn,k_x}$, where $\omega_j = A_jk_x^2 + \Omega_j$ stems from the intralayer exchange interaction and the effective static field experienced by each sublattice \cite{SM}. Here, the coefficient $A_j$ is defined as $A_j = 2A^\mathrm{ex}_j \gamma / M_{\mathrm{s}j}$, with the exchange stiffness constant $A^\mathrm{ex}_j$, electron's gyromagnetic ratio $\gamma$, and the saturated magnetization $M_{\mathrm{s}j}$. $\Omega_j = \gamma\mu_{0}H_{j}$ is the Larmor precession frequency, with vacuum permeability $\mu_0$ and the effective static field $H_j$. The sign of $H_j$ is defined for each individual layer such that a positive field is parallel to its local equilibrium moment orientation. $H_j$ includes two parts: one is the static interfacial exchange field $H^\mathrm{ex}_j$, which is always positive, and the other is the external field $H$, which is positive (negative) for the sublattice with higher (lower) saturated magnetization. Due to the dipolar interaction, magnons with the same wave vector in neighboring layers will get coupled. As previous theories and experiments reveal \cite{Yu2019, Chen2019, Ishibashi2020, Han2021}, one important feature associated with the dipolar field from propagating magnons is the chirality -- whether it emerges on the top or bottom side of the thin film depends on the sign of $k_x$ and the equilibrium moment orientation, as illustrated in Figs. \hyperref[fig:1]{1(b)} and \hyperref[fig:1]{1(c)}. Therefore, the interlayer dipolar coupling strength within and outside a unit cell are not equal to each other. The interaction Hamiltonian for the multilayers can be written as $\hat{\mathcal{H}}_\mathrm{int} = \sum_{k_x, n} (\Delta_\mathrm{S} b_{1n,k_x}^\dagger b_{2n,k_x} + \Delta_\mathrm{D} b_{1n,k_x}^\dagger b_{2,n-1,k_x} + \mathrm{H.c.})$, with the intracell (intercell) coupling strength $\Delta_\mathrm{S(D)}$. Since the interlayer dipolar interaction between magnons decays exponentially as the distance of two considered layers increases \cite{SM}, we only consider the dipolar interaction between nearest neighboring layers throughout the work. When $|k_x|$ is not large compared with the Brillouin zone (BZ) boundary, $\Delta_\mathrm{S(D)} = B \cdot (|k_x| \pm k_x)$, with $B = \gamma \mu_0 N_{1} N_{2} \sqrt{M_\mathrm{s1}d_1M_\mathrm{s2}d_2} / 2$ and $N_j = (1 - e^{-|k_x|d_j}) / (|k_x|d_j)$ \cite{SM}. In sharp contrast to the standard analytic expressions that describe local interactions in electronic systems, $\Delta_\mathrm{S(D)}$ here contains a non-analytic function of $|k_x| \pm k_x$, which is originated from the long-range nature of dipolar fields.

Combining $\hat{\mathcal{H}}_0$ and $\hat{\mathcal{H}}_\mathrm{int}$, adopting the periodic boundary condition along the $z$ axis, and implementing Fourier transformation, i.e., $b_{jn,k_x} = (1 / \sqrt{N}) \sum_{k_z} \beta_{j,\bm{k}} e^{ik_znd}$ with $\bm{k} = k_x \hat{x} + k_z \hat{z}$, we get the bulk magnonic Hamiltonian for the multilayers on the basis of $\bm{\beta}_{\bm{k}} = [\beta_{1,\bm{k}}, \beta_{2,\bm{k}}]^T$:
\begin{equation}
\begin{gathered}
\mathcal{\hat{H}}_\mathrm{bulk} = \sum_{\bm{k}}\bm{\beta}_{\bm{k}}^{\dagger} \bm{H}_\mathrm{bulk}(\bm{k})\bm{\beta}_{\bm{k}}, \\
\bm{H}_\mathrm{bulk}(\bm{k}) = \begin{bmatrix}
\omega_{1} & \Delta_\mathrm{S} + \Delta_\mathrm{D} e^{-ik_zd} \\
\Delta_\mathrm{S} + \Delta_\mathrm{D} e^{ik_zd} & \omega_{2}
\end{bmatrix}.
\end{gathered}
\label{Bulk Hamiltonian}
\end{equation}

\noindent Eq. \eqref{Bulk Hamiltonian} coincides with the expression of the celebrated one-dimensional Su-Schrieffer-Heeger (1D SSH) model \cite{Su1979}, except that our model applies to a two-dimensional (2D) case with both $\omega_{j}$ and $\Delta_\mathrm{S(D)}$ being functions of $k_x$. In the SSH model, the existence of surface/edge states is determined by the relative magnitude of $\Delta_\mathrm{S}$ and $\Delta_\mathrm{D}$. In our case, this is further controlled by the sign of $k_x$, as one can easily verify $\Delta_\mathrm{S} > \Delta_\mathrm{D}$ for $k_x > 0$, and $\Delta_\mathrm{S} < \Delta_\mathrm{D}$ for $k_x < 0$. Solving the eigenvalue equation $\bm{H}_\mathrm{bulk}\ket{\chi} = \omega\ket{\chi}$, we can get the higher (lower) eigenfrequency $\omega_{\pm} = (\omega_1 + \omega_2) / 2 \pm \sqrt{(\omega_1 - \omega_2)^2 / 4 + 4B^2k^2_x}$, either of which has no dispersions along the $k_z$ direction, suggesting a flat band when $k_x$ is fixed. Since our model is 2D in the $xz$ plane with broken TRS, we can calculate the Chern integer as the topological invariant. Denoting $\ket{\chi_-}$ as the eigenstate corresponding to $\omega_{-}$, we can calculate the Chern integer $Ch_- = (2\pi)^{-1} \iint_{\mathrm{BZ}} \Omega_{-}(\bm{k}) dk_xdk_z$ for the lower band, where $\Omega_{-}(\bm{k}) = i \partial_{k_x} \bra{\chi_{-}} \partial_{k_z} \ket{\chi_{-}} - i \partial_{k_z} \bra{\chi_{-}} \partial_{k_x} \ket{\chi_{-}}$ is the Berry curvature. The Chern integer is evaluated to be
\begin{equation} 
\begin{aligned}
Ch_- = \left\{ \begin{aligned}
&1 \quad (A_1 > A_2, H_1 < H_2)\\
-&1 \quad (A_1 < A_2, H_1 > H_2)\\
&0 \quad (\mathrm{otherwise})
\end{aligned} \right.
\end{aligned},
\label{Chern integer}
\end{equation}

\noindent suggesting that the bulk bands possess non-zero Chern integers when $(A_1 - A_2) \cdot (H_1 - H_2) < 0$.

\begin{figure}[t]
	\centering
	\includegraphics[width=1\columnwidth]{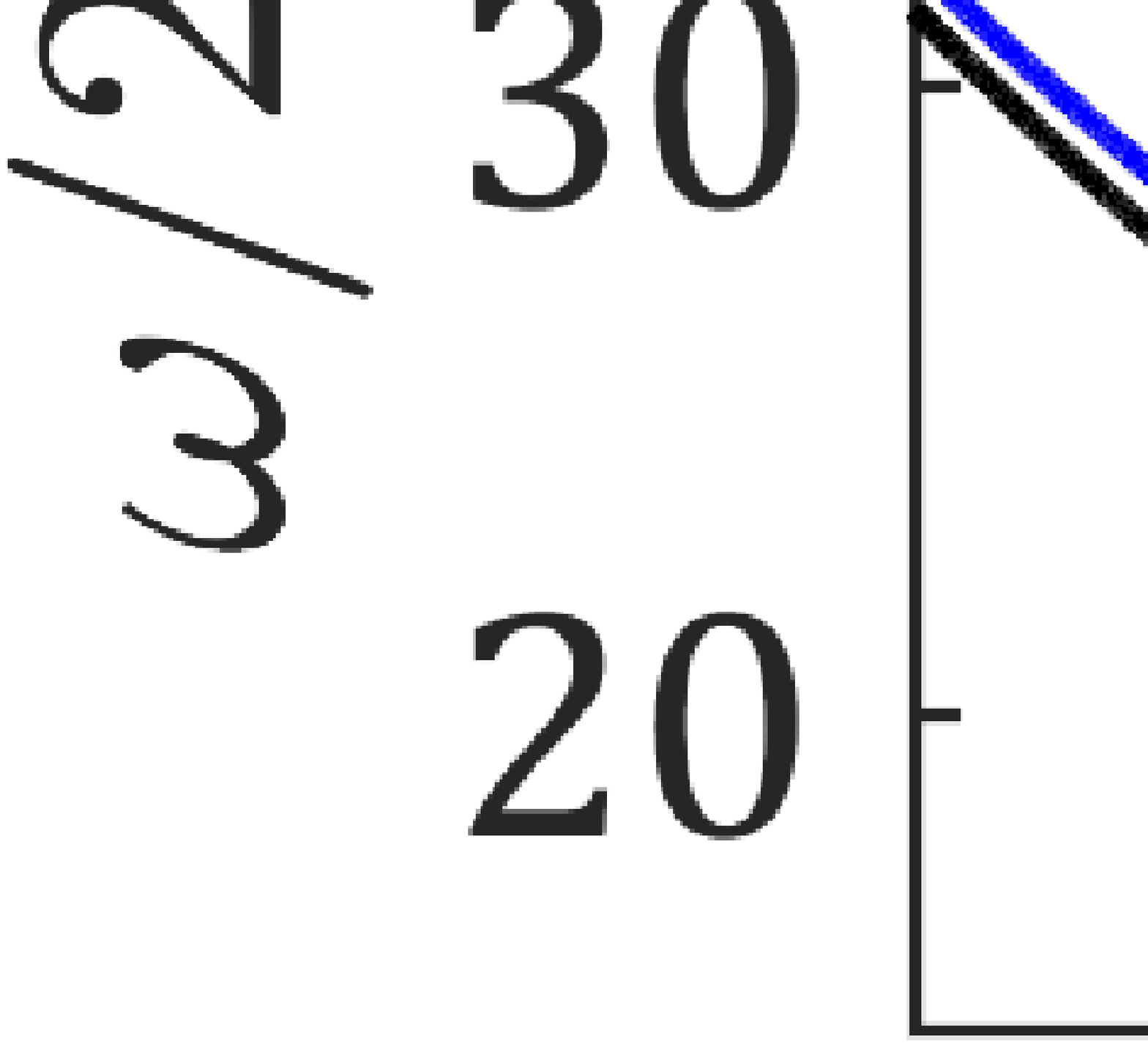}
	\caption{(a) to (c) Evolution of the magnonic band structure with a varying external field $H$. The black lines correspond to bulk states, and the red (blue) line corresponds to the surface state strictly localized in the bottom (top) layer. (a) and (c) correspond to the topologically non-trivial and trivial phases, respectively. (b) is when the topological transition happens. (d) Coupled wire construction in the magnetic multilayers with $2N$ total layers. Due to the chirality of the interlayer dipolar interaction, right-moving magnons get coupled (red arrows) within the same cell, while left-moving magnons get coupled (green arrows) between neighboring cells. Two surface states are left uncoupled. The symbols of $\otimes$ and $\odot$ on the surface states show the orientations of carried spins. (e) Coupled wire construction in the multilayers with $2N - 1$ total layers.}
	\label{fig:2}
\end{figure}

The surface states can be explicitly obtained by considering multilayers with open boundary condition along the $z$ direction. The magnonic Hamiltonian for the $N$-cell multilayers is $\hat{\mathcal{H}} = \hat{\mathcal{H}}_0 + \hat{\mathcal{H}}_\mathrm{int} = \sum_{k_x}\bm{b}_{k_x}^{\dagger} \bm{H}(k_x)\bm{b}_{k_x}$, where $\bm{b}_{k_x} = [b_{11,k_x}, b_{21,k_x}, ..., b_{1N,k_x}, b_{2N,k_x}]^T$ and $\bm{H}(k_x)$ is a $2N \times 2N$ matrix. Solving the eigenvalue equation $\bm{H}\ket{\chi} = \omega\ket{\chi}$, we can get $2N$ eigenfrequencies and corresponding eigenstates. When $k_x < 0$, corresponding to $\Delta_\mathrm{S} = 0$ and $\Delta_\mathrm{D} > 0$, two surface states emerge, which are separately localized in the $1$st layer of the $1$st cell (bottom layer) with $\omega^\mathrm{1e} = \omega_1$, $\ket{\chi^\mathrm{1e}} = [1, 0, 0, ..., 0]^T$, and in the $2$nd layer of the $N$th cell (top layer) with $\omega^\mathrm{2e} = \omega_2$, $\ket{\chi^\mathrm{2e}} = [0, 0, ..., 0, 1]^T$. We note that both of them are strictly localized on the surfaces without decay into the bulk. Figs. \hyperref[fig:2]{2(a)} to \hyperref[fig:2]{2(c)} show the evolution of the magnonic band structure, when the strength of applied external magnetic field $H$ varies. In this calculation, we consider an example consisting of alternating $10 \ \mathrm{nm}$ yttrium iron garnet (YIG, $1$st sublattice) and $10 \ \mathrm{nm}$ permalloy (Py, $2$nd sublattice) thin films, with reported material parameters of $A^{\mathrm{ex}}_1 = 3.7 \times 10^{-12} \ \mathrm{J/m}$, $A^{\mathrm{ex}}_2 = 8.7 \times 10^{-12} \ \mathrm{J/m}$, $M_{\mathrm{s1}} = 1.4 \times 10^5 \ \mathrm{A/m}$, and $M_{\mathrm{s2}} = 7.4 \times 10^5 \ \mathrm{A/m}$ \cite{Klingler2014, Langer2016}. An external magnetic field is assumed along the $-y$ direction. The antiferromagnetic exchange coupling constant at the YIG/Py interface is taken to be $J = - 8.6 \times 10^4 \ \mathrm{J/m^2}$ \cite{Fan2020}, which gives rise to the static interfacial exchange field $H^{\mathrm{ex}}_j = 2|J| / (\mu_0 M_{\mathrm{s}j}d_j)$ experienced by each sublattice, further leading to the total effective static field $H_j = H^{\mathrm{ex}}_j \mp H$. As the Chern integer calculation shows, the topology of the magnonic band structure depends on the sign of $(A_1 - A_2) \cdot (H_1 - H_2)$, which can be further controlled by tuning the relative magnitude of $H_1$ and $H_2$ through $H$, given that $A_1 - A_2 > 0$ is fixed. In Fig. \hyperref[fig:2]{2(a)}, under $H = 5 \times 10^5 \ \mathrm{A/m}$, corresponding to $H_1 < H_2$, the bulk bands (black lines) are inverted around $k_x$ = 0 and two surface bands (red and blue lines) emerge in the $k_x < 0$ half-space, which cross each other and form a degenerate point, i.e., a tilted Dirac cone, without pairing at its time-reversal point due to broken TRS. When $H$ is reduced such that $H_1 = H_2$, a topological transition happens [Fig. \hyperref[fig:2]{2(b)}], where the bulk bands become degenerate at $k_x = 0$, corresponding to a gap closing. With further decreasing $H$ such that $H_1 > H_2$ [Fig. \hyperref[fig:2]{2(c)}], there is no bulk band inversion and the surface bands do not cross each other in the $k_x < 0$ half-space, representing a topologically trivial system.

The formation of ultra-localized surface states in the magnetic multilayers can be further understood as a magnonic implementation of the coupled wire model that has been widely investigated in the circumstance of quantum Hall effect (QHE) in electronic systems \cite{Kane2002, Meng2015, Kane2018, Wu2019}. We begin by regarding the multilayers as an array of $2N$ non-interacting one-dimensional wires, with single-particle magnonic dispersion relations $\omega_{1(2)}(k_x)$, which cross each other under the condition of $(A_1 - A_2) \cdot (H_1 - H_2) < 0$. Due to the interlayer dipolar interaction, magnons around the band-crossing points in neighboring layers get coupled, as shown in Fig. \hyperref[fig:2]{2(d)}. As discussed earlier, depending on the sign of $k_x$, either $\Delta_\mathrm{S}$ or $\Delta_\mathrm{D}$ reduces to zero. For right-moving magnons ($k_x > 0$), $\Delta_\mathrm{D} = 0$ and $\Delta_\mathrm{S} > 0$, the coupling only happens within the same cell [red arrows in Fig. \hyperref[fig:2]{2(d)}], while for left-moving magnons ($k_x < 0$), $\Delta_\mathrm{S} = 0$ and $\Delta_\mathrm{D} > 0$, the coupling only happens between neighboring cells (green arrows). As a result, the left-moving magnons in the bottom layer and the left-moving magnons in the top layer are left uncoupled and form a pair of surface states. Considering that two surface layers possess opposite equilibrium moments, the surface magnonic states would carry spin currents with opposite directions. This kind of chiral surface spin currents still exist even if the number of layers is odd, as shown in Fig. \hyperref[fig:2]{2(e)}. In this case, the surface states have opposite velocities but same equilibrium moment orientations, hence still carrying opposite spin currents. The formation of surface magnonic states and chiral spin currents in the magnetic multilayers is therefore similar to the realization of QHE in electronic systems, where the left- and right-moving electrons in neighboring wires are coupled together due to interchannel scattering.

Until now, we have demonstrated tunable magnonic Chern bands and ultra-localized surface states carrying chiral spin currents in the magnetic multilayers under a few simplified assumptions, including: 1) the magnon propagation is confined in the $x$ direction, which is orthogonal to the equilibrium moment orientations, 2) we neglect the intralayer dipolar interaction and dynamic interfacial exchange interaction. In the following, we extend our discussion to a generic case where magnons can propagate within the whole $xy$ plane in each individual layer, i.e., with in-plane momenta $\bm{k_\parallel} = k_x \hat{x} + k_y \hat{y}$, and both the intralayer dipolar interaction and dynamic interfacial exchange interaction are included. With the periodic boundary condition along the $z$ axis, the bulk magnonic Hamiltonian for the multilayers in Fig. \hyperref[fig:1]{1(a)} can be expressed in the Bogoliubov-de Gennes formalism:
\begin{equation}
\begin{gathered}
\mathcal{\hat{H}}^\prime_\mathrm{bulk} = \frac{1}{2}\sum_{\bm{k}}\begin{bmatrix} \bm{\beta}_{\bm{k}}^{\dagger} & \bm{\beta}_{-\bm{k}} \end{bmatrix} \bm{H}^\prime_\mathrm{bulk}(\bm{k})\begin{bmatrix} \bm{\beta}_{\bm{k}} \\ \bm{\beta}_{-\bm{k}}^{\dagger} \end{bmatrix}, \\
\bm{H}^\prime_\mathrm{bulk}(\bm{k}) = \begin{bmatrix}
\bm{h}(\bm{k}) & \bm{b}(\bm{k}) \\
\bm{b}^{*}(-\bm{k}) & \bm{h}^{*}(-\bm{k}) \end{bmatrix},
\end{gathered}
\label{new bulk H}
\end{equation}

\noindent with $\bm{k} = \bm{k}_\parallel + k_z \hat{z}$. The diagonal block $\bm{h}(\bm{k})$ in Eq. \eqref{new bulk H} has the form of
\begin{equation}
\begin{gathered}
	\bm{h}(\bm{k}) = \begin{bmatrix}
	\omega_{1}^\prime & \Delta_\mathrm{S}^\prime + \Delta_\mathrm{D}^\prime e^{-ik_zd} \\
	\Delta_\mathrm{S}^\prime + \Delta_\mathrm{D}^\prime e^{ik_zd} & \omega_{2}^\prime
	\end{bmatrix},
\end{gathered}
\end{equation}

\noindent with
\begin{equation}
\begin{gathered}
\omega_j^\prime = A_j\bm{k}_\parallel^2 + \Omega_j + \frac{\gamma\mu_0M_{\mathrm{s}j}}{2}\left[(1 - N_{j}^\prime)\frac{k_x^2}{k_\parallel^2} + N_{j}^\prime + \frac{2}{3}\right], \\
\Delta_{\mathrm{S(D)}}^\prime = \frac{B^\prime}{2} \cdot \left(k_\parallel + \frac{k_x^2}{k_\parallel} \pm 2k_x\right),
\end{gathered}
\label{new coupling term}
\end{equation}

\begin{figure}[t]
	\includegraphics[width=1\columnwidth]{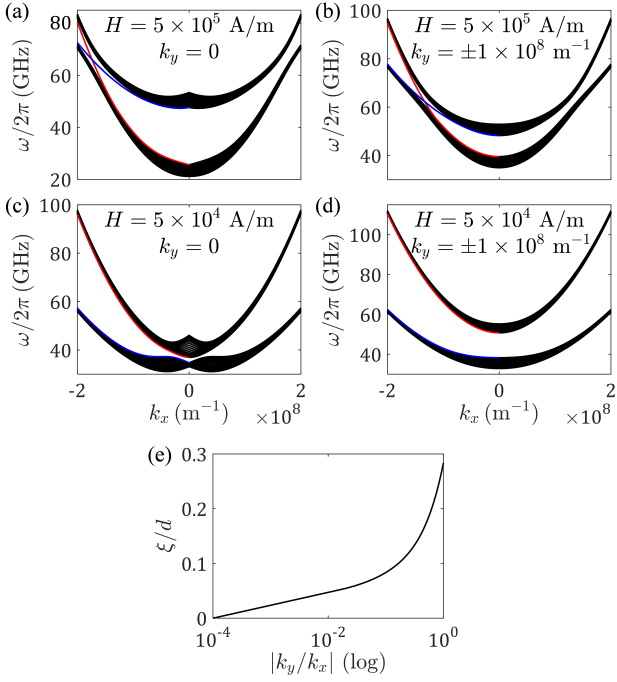}
	\caption{(a) to (d) Magnonic band structure for the 3D multilayers when all the interactions are included. (a) and (b) correspond to the topologically non-trivial phase with $H = 5 \times 10^5 \ \mathrm{A/m}$, while (c) and (d) correspond to the trivial phase with $H = 5 \times 10^4 \ \mathrm{A/m}$. $k_y$ is fixed at $0$ for (a) and (c), and $k_y$ is fixed at $\pm 1.0 \times 10^8 \ \mathrm{m^{-1}}$ for (b) and (d). (e) Decay length $\xi$ of surface states as a function of $|k_y / k_x|$.}
	\label{fig:3}
\end{figure}

\noindent where $N_j^\prime$ and $B^\prime$ can be obtained by replacing $|k_x|$ with $|\bm{k}_\parallel|$ in $N_j$ and $B$. The off-diagonal block $\bm{b}(\bm{k})$ in Eq. \eqref{new bulk H} has the form of
\begin{equation}
\begin{gathered}
\bm{b}(\bm{k}) = \begin{bmatrix}
\delta_{1} & \delta^{\prime} \cdot (1 + e^{-ik_zd}) \\
\delta^{\prime} \cdot (1 + e^{ik_zd}) & \delta_{2}
\end{bmatrix},
\end{gathered}
\end{equation}

\noindent with
\begin{equation}
\begin{gathered}
\delta_{j} = \frac{\gamma\mu_0M_{\mathrm{s}j}}{2} \left[(1 - N_{j}^\prime)\frac{k_x^2}{k_\parallel^2} - N_{j}^\prime\right], \\
\delta^{\prime} = \frac{\gamma\mu_0}{2} \prod_{j = 1, 2}\left[f_j^\prime \cdot \left( k_\parallel - \frac{k_x^2}{k_\parallel} \right) + H^\mathrm{ex}_j\right]^{1/2},
\end{gathered}
\label{elements in b}
\end{equation}

\noindent where $f_{1(2)}^\prime = M_{\mathrm{s}2(1)} d_{2(1)} N_1^\prime N_2^\prime / 2$. The derivation for Eqs. \eqref{new bulk H} to \eqref{elements in b} is presented in \cite{SM}. Here, we note that $\beta_{\bm k}$ ($\beta_{\bm k}^\dagger$) in Eq. \eqref{new bulk H} corresponds to the annihilation (creation) operator for circularly polarized right-handed magnon modes with positive frequencies, or equivalently, particles, while $\beta_{-{\bm k}}^{\dagger}$ ($\beta_{-{\bm k}}$) represents the annihilation (creation) operator for left-handed magnon modes with negative frequencies, or equivalently, holes. The intralayer dipolar interaction mixes magnon modes with opposite handedness, i.e., particles and holes, within the same layer, and leads to non-zero $\delta_{j}$ in $\bm{b}(\bm{k})$. Meanwhile, the dynamic interfacial exchange interaction and the interlayer dipolar interaction under finite $k_y$ give rise to the terms proportional to $\delta^\prime$ in $\bm{b}(\bm{k})$, coupling magnon modes with opposite handedness in neighboring layers. For the usual cases with large enough effective static field $H_j$ that we are interested in, the difference between frequencies of right- and left-handed magnon modes is large, so $\bm{b}(\bm{k})$ in Eq. \eqref{new bulk H} plays a role as a weak perturbation, which does not lead to extra band inversions. Therefore approximately, the existence of surface states is still governed by the relative magnitude of $\Delta_\mathrm{S}^\prime$ and $\Delta_\mathrm{D}^\prime$ in $\bm{h}(\bm{k})$, which is further controlled by the sign of $k_x$ according to Eq. \eqref{new coupling term}.

With open boundary condition along the $z$ direction, the magnonic band structure can be solved through numerical methods based on the Cholesky decomposition \cite{Colpa1978}. Using the same parameters as in Figs. \hyperref[fig:2]{2(a)} to \hyperref[fig:2]{2(c)}, we plot the magnonic band structure in Figs. \hyperref[fig:3]{3(a)} to \hyperref[fig:3]{3(d)} with fixed $k_y$ for two different $H$, when all the interactions are included. We see that after extending to a generic three-dimensional (3D) case, the surface states still exist in the $k_x < 0$ half-space. When $H$ remains the same and $k_y$ varies, the band structure keeps its topological properties, either with non-trivial [Figs. \hyperref[fig:3]{3(a)} and \hyperref[fig:3]{3(b)}] or trivial [Figs. \hyperref[fig:3]{3(c)} and \hyperref[fig:3]{3(d)}] surface states, suggesting that the 3D multilayers can be regarded as the magnonic analog of stacked 2D Chern `insulators' along the $y$ direction \cite{Fu2007}. Similar to the 2D simplified case discussed earlier, the topological properties are still tunable through $H$, i.e., the surface states can still be toggled between non-trivial and trivial phases, as illustrated by the comparison of Figs. \hyperref[fig:3]{3(b)} and \hyperref[fig:3]{3(d)}. The subtle difference is that with finite $k_y$, $\Delta_\mathrm{S}^\prime$ no longer vanishes for $k_x < 0$, and the surface states therefore extend into the bulk with an exponential decay in magnitude, $\ket{\chi^\mathrm{1e}}(z) \sim e^{-z/\xi}$ and $\ket{\chi^\mathrm{2e}}(z) \sim e^{-(z - Nd)/\xi}$, where $\xi = d / \rm{log}(\Delta_\mathrm{D}^\prime/\Delta_\mathrm{S}^\prime)$ is the decay length. In Fig. \hyperref[fig:3]{3(e)}, we plot $\xi$ as a function of $|k_y / k_x|$, and see that $\xi \ll d$ for $k_y < k_x$, indicating the surfaces states are still highly localized.  

In summary, we study the magnonic band structure and corresponding topological properties in antiparallelly aligned magnetic multilayers. We demonstrate that the long-range, chiral interlayer dipolar interaction between propagating magnons correlates the sublattice and momentum degrees of freedom and breaks TRS. It further generates bulk bands with non-zero Chern integers and ultra-localized surface states carrying chiral spin currents. The topology of magnonic bands can be switched between non-trivial and trivial through an external field. We also reveal that the dipolarly coupled magnetic multilayers represents a magnonic implementation of the coupled wire model. We believe our study provides an easy-to-implement system for realizing topologically protected magnonic surface states and low-dissipation spin transport in a tunable manner, and will therefore benefit the development of modern spintronics.

\begin{acknowledgments}
This work is supported by AFOSR under award FA9550-19-1-0048, and National Science Foundation under award DMR-2104912.
\end{acknowledgments}

\bibliography{main_text}

\begin{thebibliography}{40}%
\makeatletter
\providecommand \@ifxundefined [1]{%
 \@ifx{#1\undefined}
}%
\providecommand \@ifnum [1]{%
 \ifnum #1\expandafter \@firstoftwo
 \else \expandafter \@secondoftwo
 \fi
}%
\providecommand \@ifx [1]{%
 \ifx #1\expandafter \@firstoftwo
 \else \expandafter \@secondoftwo
 \fi
}%
\providecommand \natexlab [1]{#1}%
\providecommand \enquote  [1]{``#1''}%
\providecommand \bibnamefont  [1]{#1}%
\providecommand \bibfnamefont [1]{#1}%
\providecommand \citenamefont [1]{#1}%
\providecommand \href@noop [0]{\@secondoftwo}%
\providecommand \href [0]{\begingroup \@sanitize@url \@href}%
\providecommand \@href[1]{\@@startlink{#1}\@@href}%
\providecommand \@@href[1]{\endgroup#1\@@endlink}%
\providecommand \@sanitize@url [0]{\catcode `\\12\catcode `\$12\catcode
  `\&12\catcode `\#12\catcode `\^12\catcode `\_12\catcode `\%12\relax}%
\providecommand \@@startlink[1]{}%
\providecommand \@@endlink[0]{}%
\providecommand \url  [0]{\begingroup\@sanitize@url \@url }%
\providecommand \@url [1]{\endgroup\@href {#1}{\urlprefix }}%
\providecommand \urlprefix  [0]{URL }%
\providecommand \Eprint [0]{\href }%
\providecommand \doibase [0]{https://doi.org/}%
\providecommand \selectlanguage [0]{\@gobble}%
\providecommand \bibinfo  [0]{\@secondoftwo}%
\providecommand \bibfield  [0]{\@secondoftwo}%
\providecommand \translation [1]{[#1]}%
\providecommand \BibitemOpen [0]{}%
\providecommand \bibitemStop [0]{}%
\providecommand \bibitemNoStop [0]{.\EOS\space}%
\providecommand \EOS [0]{\spacefactor3000\relax}%
\providecommand \BibitemShut  [1]{\csname bibitem#1\endcsname}%
\let\auto@bib@innerbib\@empty
\bibitem [{\citenamefont {Hasan}\ and\ \citenamefont {Kane}(2010)}]{Hasan2010}%
  \BibitemOpen
  \bibfield  {author} {\bibinfo {author} {\bibfnamefont {M.~Z.}\ \bibnamefont
  {Hasan}}\ and\ \bibinfo {author} {\bibfnamefont {C.~L.}\ \bibnamefont
  {Kane}},\ }\bibfield  {title} {\bibinfo {title} {Colloquium: Topological
  insulators},\ }\href {https://doi.org/10.1103/RevModPhys.82.3045} {\bibfield
  {journal} {\bibinfo  {journal} {Rev. Mod. Phys.}\ }\textbf {\bibinfo {volume}
  {82}},\ \bibinfo {pages} {3045} (\bibinfo {year} {2010})}\BibitemShut
  {NoStop}%
\bibitem [{\citenamefont {Qi}\ and\ \citenamefont {Zhang}(2011)}]{Qi2011}%
  \BibitemOpen
  \bibfield  {author} {\bibinfo {author} {\bibfnamefont {X.-L.}\ \bibnamefont
  {Qi}}\ and\ \bibinfo {author} {\bibfnamefont {S.-C.}\ \bibnamefont {Zhang}},\
  }\bibfield  {title} {\bibinfo {title} {Topological insulators and
  superconductors},\ }\href {https://doi.org/10.1103/RevModPhys.83.1057}
  {\bibfield  {journal} {\bibinfo  {journal} {Rev. Mod. Phys.}\ }\textbf
  {\bibinfo {volume} {83}},\ \bibinfo {pages} {1057} (\bibinfo {year}
  {2011})}\BibitemShut {NoStop}%
\bibitem [{\citenamefont {Ando}\ and\ \citenamefont {Fu}(2015)}]{Ando2015}%
  \BibitemOpen
  \bibfield  {author} {\bibinfo {author} {\bibfnamefont {Y.}~\bibnamefont
  {Ando}}\ and\ \bibinfo {author} {\bibfnamefont {L.}~\bibnamefont {Fu}},\
  }\bibfield  {title} {\bibinfo {title} {Topological crystalline insulators and
  topological superconductors: From concepts to materials},\ }\href
  {https://doi.org/10.1146/annurev-conmatphys-031214-014501} {\bibfield
  {journal} {\bibinfo  {journal} {Annu. Rev. Condens. Matter Phys.}\ }\textbf
  {\bibinfo {volume} {6}},\ \bibinfo {pages} {361} (\bibinfo {year}
  {2015})}\BibitemShut {NoStop}%
\bibitem [{\citenamefont {Sato}\ and\ \citenamefont {Ando}(2017)}]{Sato2017}%
  \BibitemOpen
  \bibfield  {author} {\bibinfo {author} {\bibfnamefont {M.}~\bibnamefont
  {Sato}}\ and\ \bibinfo {author} {\bibfnamefont {Y.}~\bibnamefont {Ando}},\
  }\bibfield  {title} {\bibinfo {title} {Topological superconductors: a
  review},\ }\href {https://doi.org/10.1088/1361-6633/aa6ac7} {\bibfield
  {journal} {\bibinfo  {journal} {Rep. Prog. Phys.}\ }\textbf {\bibinfo
  {volume} {80}},\ \bibinfo {pages} {076501} (\bibinfo {year}
  {2017})}\BibitemShut {NoStop}%
\bibitem [{\citenamefont {Armitage}\ \emph {et~al.}(2018)\citenamefont
  {Armitage}, \citenamefont {Mele},\ and\ \citenamefont
  {Vishwanath}}]{Armitage2018}%
  \BibitemOpen
  \bibfield  {author} {\bibinfo {author} {\bibfnamefont {N.~P.}\ \bibnamefont
  {Armitage}}, \bibinfo {author} {\bibfnamefont {E.~J.}\ \bibnamefont {Mele}},\
  and\ \bibinfo {author} {\bibfnamefont {A.}~\bibnamefont {Vishwanath}},\
  }\bibfield  {title} {\bibinfo {title} {Weyl and dirac semimetals in
  three-dimensional solids},\ }\href
  {https://doi.org/10.1103/RevModPhys.90.015001} {\bibfield  {journal}
  {\bibinfo  {journal} {Rev. Mod. Phys.}\ }\textbf {\bibinfo {volume} {90}},\
  \bibinfo {pages} {015001} (\bibinfo {year} {2018})}\BibitemShut {NoStop}%
\bibitem [{\citenamefont {Wang}\ \emph {et~al.}(2015)\citenamefont {Wang},
  \citenamefont {Lu},\ and\ \citenamefont {Bertoldi}}]{Wang2015}%
  \BibitemOpen
  \bibfield  {author} {\bibinfo {author} {\bibfnamefont {P.}~\bibnamefont
  {Wang}}, \bibinfo {author} {\bibfnamefont {L.}~\bibnamefont {Lu}},\ and\
  \bibinfo {author} {\bibfnamefont {K.}~\bibnamefont {Bertoldi}},\ }\bibfield
  {title} {\bibinfo {title} {Topological phononic crystals with one-way elastic
  edge waves},\ }\href {https://doi.org/10.1103/PhysRevLett.115.104302}
  {\bibfield  {journal} {\bibinfo  {journal} {Phys. Rev. Lett.}\ }\textbf
  {\bibinfo {volume} {115}},\ \bibinfo {pages} {104302} (\bibinfo {year}
  {2015})}\BibitemShut {NoStop}%
\bibitem [{\citenamefont {He}\ \emph {et~al.}(2018)\citenamefont {He},
  \citenamefont {Qiu}, \citenamefont {Ye}, \citenamefont {Cai}, \citenamefont
  {Fan}, \citenamefont {Ke}, \citenamefont {Zhang},\ and\ \citenamefont
  {Liu}}]{He2018}%
  \BibitemOpen
  \bibfield  {author} {\bibinfo {author} {\bibfnamefont {H.}~\bibnamefont
  {He}}, \bibinfo {author} {\bibfnamefont {C.}~\bibnamefont {Qiu}}, \bibinfo
  {author} {\bibfnamefont {L.}~\bibnamefont {Ye}}, \bibinfo {author}
  {\bibfnamefont {X.}~\bibnamefont {Cai}}, \bibinfo {author} {\bibfnamefont
  {X.}~\bibnamefont {Fan}}, \bibinfo {author} {\bibfnamefont {M.}~\bibnamefont
  {Ke}}, \bibinfo {author} {\bibfnamefont {F.}~\bibnamefont {Zhang}},\ and\
  \bibinfo {author} {\bibfnamefont {Z.}~\bibnamefont {Liu}},\ }\bibfield
  {title} {\bibinfo {title} {Topological negative refraction of surface
  acoustic waves in a weyl phononic crystal},\ }\href
  {https://doi.org/10.1038/s41586-018-0367-9} {\bibfield  {journal} {\bibinfo
  {journal} {Nature (London)}\ }\textbf {\bibinfo {volume} {560}},\ \bibinfo
  {pages} {61} (\bibinfo {year} {2018})}\BibitemShut {NoStop}%
\bibitem [{\citenamefont {Wu}\ and\ \citenamefont {Hu}(2015)}]{Wu2015}%
  \BibitemOpen
  \bibfield  {author} {\bibinfo {author} {\bibfnamefont {L.-H.}\ \bibnamefont
  {Wu}}\ and\ \bibinfo {author} {\bibfnamefont {X.}~\bibnamefont {Hu}},\
  }\bibfield  {title} {\bibinfo {title} {Scheme for achieving a topological
  photonic crystal by using dielectric material},\ }\href
  {https://doi.org/10.1103/PhysRevLett.114.223901} {\bibfield  {journal}
  {\bibinfo  {journal} {Phys. Rev. Lett.}\ }\textbf {\bibinfo {volume} {114}},\
  \bibinfo {pages} {223901} (\bibinfo {year} {2015})}\BibitemShut {NoStop}%
\bibitem [{\citenamefont {Lu}\ \emph {et~al.}(2016)\citenamefont {Lu},
  \citenamefont {Fang}, \citenamefont {Fu}, \citenamefont {Johnson},
  \citenamefont {Joannopoulos},\ and\ \citenamefont
  {Solja{\v{c}}i{\'c}}}]{Lu2016}%
  \BibitemOpen
  \bibfield  {author} {\bibinfo {author} {\bibfnamefont {L.}~\bibnamefont
  {Lu}}, \bibinfo {author} {\bibfnamefont {C.}~\bibnamefont {Fang}}, \bibinfo
  {author} {\bibfnamefont {L.}~\bibnamefont {Fu}}, \bibinfo {author}
  {\bibfnamefont {S.~G.}\ \bibnamefont {Johnson}}, \bibinfo {author}
  {\bibfnamefont {J.~D.}\ \bibnamefont {Joannopoulos}},\ and\ \bibinfo {author}
  {\bibfnamefont {M.}~\bibnamefont {Solja{\v{c}}i{\'c}}},\ }\bibfield  {title}
  {\bibinfo {title} {Symmetry-protected topological photonic crystal in three
  dimensions},\ }\href {https://doi.org/10.1038/nphys3611} {\bibfield
  {journal} {\bibinfo  {journal} {Nat. Phys.}\ }\textbf {\bibinfo {volume}
  {12}},\ \bibinfo {pages} {337} (\bibinfo {year} {2016})}\BibitemShut
  {NoStop}%
\bibitem [{\citenamefont {Kajiwara}\ \emph {et~al.}(2010)\citenamefont
  {Kajiwara}, \citenamefont {Harii}, \citenamefont {Takahashi}, \citenamefont
  {Ohe}, \citenamefont {Uchida}, \citenamefont {Mizuguchi}, \citenamefont
  {Umezawa}, \citenamefont {Kawai}, \citenamefont {Ando}, \citenamefont
  {Takanashi}, \citenamefont {Maekawa},\ and\ \citenamefont
  {Saitoh}}]{Kajiwara2010}%
  \BibitemOpen
  \bibfield  {author} {\bibinfo {author} {\bibfnamefont {Y.}~\bibnamefont
  {Kajiwara}}, \bibinfo {author} {\bibfnamefont {K.}~\bibnamefont {Harii}},
  \bibinfo {author} {\bibfnamefont {S.}~\bibnamefont {Takahashi}}, \bibinfo
  {author} {\bibfnamefont {J.}~\bibnamefont {Ohe}}, \bibinfo {author}
  {\bibfnamefont {K.}~\bibnamefont {Uchida}}, \bibinfo {author} {\bibfnamefont
  {M.}~\bibnamefont {Mizuguchi}}, \bibinfo {author} {\bibfnamefont
  {H.}~\bibnamefont {Umezawa}}, \bibinfo {author} {\bibfnamefont
  {H.}~\bibnamefont {Kawai}}, \bibinfo {author} {\bibfnamefont
  {K.}~\bibnamefont {Ando}}, \bibinfo {author} {\bibfnamefont {K.}~\bibnamefont
  {Takanashi}}, \bibinfo {author} {\bibfnamefont {S.}~\bibnamefont {Maekawa}},\
  and\ \bibinfo {author} {\bibfnamefont {E.}~\bibnamefont {Saitoh}},\
  }\bibfield  {title} {\bibinfo {title} {Transmission of electrical signals by
  spin-wave interconversion in a magnetic insulator},\ }\href
  {https://doi.org/10.1038/nature08876} {\bibfield  {journal} {\bibinfo
  {journal} {Nature (London)}\ }\textbf {\bibinfo {volume} {464}},\ \bibinfo
  {pages} {262} (\bibinfo {year} {2010})}\BibitemShut {NoStop}%
\bibitem [{\citenamefont {Cornelissen}\ \emph {et~al.}(2015)\citenamefont
  {Cornelissen}, \citenamefont {Liu}, \citenamefont {Duine}, \citenamefont
  {Youssef},\ and\ \citenamefont {van Wees}}]{Cornelissen2015}%
  \BibitemOpen
  \bibfield  {author} {\bibinfo {author} {\bibfnamefont {L.~J.}\ \bibnamefont
  {Cornelissen}}, \bibinfo {author} {\bibfnamefont {J.}~\bibnamefont {Liu}},
  \bibinfo {author} {\bibfnamefont {R.~A.}\ \bibnamefont {Duine}}, \bibinfo
  {author} {\bibfnamefont {J.~B.}\ \bibnamefont {Youssef}},\ and\ \bibinfo
  {author} {\bibfnamefont {B.~J.}\ \bibnamefont {van Wees}},\ }\bibfield
  {title} {\bibinfo {title} {Long-distance transport of magnon spin information
  in a magnetic insulator at room temperature},\ }\href
  {https://doi.org/10.1038/nphys3465} {\bibfield  {journal} {\bibinfo
  {journal} {Nat. Phys.}\ }\textbf {\bibinfo {volume} {11}},\ \bibinfo {pages}
  {1022} (\bibinfo {year} {2015})}\BibitemShut {NoStop}%
\bibitem [{\citenamefont {Han}\ \emph {et~al.}(2019)\citenamefont {Han},
  \citenamefont {Zhang}, \citenamefont {Hou}, \citenamefont {Siddiqui},\ and\
  \citenamefont {Liu}}]{Han2019}%
  \BibitemOpen
  \bibfield  {author} {\bibinfo {author} {\bibfnamefont {J.}~\bibnamefont
  {Han}}, \bibinfo {author} {\bibfnamefont {P.}~\bibnamefont {Zhang}}, \bibinfo
  {author} {\bibfnamefont {J.~T.}\ \bibnamefont {Hou}}, \bibinfo {author}
  {\bibfnamefont {S.~A.}\ \bibnamefont {Siddiqui}},\ and\ \bibinfo {author}
  {\bibfnamefont {L.}~\bibnamefont {Liu}},\ }\bibfield  {title} {\bibinfo
  {title} {Mutual control of coherent spin waves and magnetic domain walls in a
  magnonic device},\ }\href {https://doi.org/10.1126/science.aau2610}
  {\bibfield  {journal} {\bibinfo  {journal} {Science}\ }\textbf {\bibinfo
  {volume} {366}},\ \bibinfo {pages} {1121} (\bibinfo {year}
  {2019})}\BibitemShut {NoStop}%
\bibitem [{\citenamefont {Turner}(1960)}]{Turner1960}%
  \BibitemOpen
  \bibfield  {author} {\bibinfo {author} {\bibfnamefont {E.~H.}\ \bibnamefont
  {Turner}},\ }\bibfield  {title} {\bibinfo {title} {Interaction of phonons and
  spin waves in yttrium iron garnet},\ }\href
  {https://doi.org/10.1103/PhysRevLett.5.100} {\bibfield  {journal} {\bibinfo
  {journal} {Phys. Rev. Lett.}\ }\textbf {\bibinfo {volume} {5}},\ \bibinfo
  {pages} {100} (\bibinfo {year} {1960})}\BibitemShut {NoStop}%
\bibitem [{\citenamefont {Raquet}\ \emph {et~al.}(2002)\citenamefont {Raquet},
  \citenamefont {Viret}, \citenamefont {Sondergard}, \citenamefont {Cespedes},\
  and\ \citenamefont {Mamy}}]{Raquet2002}%
  \BibitemOpen
  \bibfield  {author} {\bibinfo {author} {\bibfnamefont {B.}~\bibnamefont
  {Raquet}}, \bibinfo {author} {\bibfnamefont {M.}~\bibnamefont {Viret}},
  \bibinfo {author} {\bibfnamefont {E.}~\bibnamefont {Sondergard}}, \bibinfo
  {author} {\bibfnamefont {O.}~\bibnamefont {Cespedes}},\ and\ \bibinfo
  {author} {\bibfnamefont {R.}~\bibnamefont {Mamy}},\ }\bibfield  {title}
  {\bibinfo {title} {Electron-magnon scattering and magnetic resistivity in
  $3d$ ferromagnets},\ }\href {https://doi.org/10.1103/PhysRevB.66.024433}
  {\bibfield  {journal} {\bibinfo  {journal} {Phys. Rev. B}\ }\textbf {\bibinfo
  {volume} {66}},\ \bibinfo {pages} {024433} (\bibinfo {year}
  {2002})}\BibitemShut {NoStop}%
\bibitem [{\citenamefont {Boona}\ and\ \citenamefont
  {Heremans}(2014)}]{Boona2014}%
  \BibitemOpen
  \bibfield  {author} {\bibinfo {author} {\bibfnamefont {S.~R.}\ \bibnamefont
  {Boona}}\ and\ \bibinfo {author} {\bibfnamefont {J.~P.}\ \bibnamefont
  {Heremans}},\ }\bibfield  {title} {\bibinfo {title} {Magnon thermal mean free
  path in yttrium iron garnet},\ }\href
  {https://doi.org/10.1103/PhysRevB.90.064421} {\bibfield  {journal} {\bibinfo
  {journal} {Phys. Rev. B}\ }\textbf {\bibinfo {volume} {90}},\ \bibinfo
  {pages} {064421} (\bibinfo {year} {2014})}\BibitemShut {NoStop}%
\bibitem [{\citenamefont {Zhang}\ \emph {et~al.}(2013)\citenamefont {Zhang},
  \citenamefont {Ren}, \citenamefont {Wang},\ and\ \citenamefont
  {Li}}]{Zhang2013}%
  \BibitemOpen
  \bibfield  {author} {\bibinfo {author} {\bibfnamefont {L.}~\bibnamefont
  {Zhang}}, \bibinfo {author} {\bibfnamefont {J.}~\bibnamefont {Ren}}, \bibinfo
  {author} {\bibfnamefont {J.-S.}\ \bibnamefont {Wang}},\ and\ \bibinfo
  {author} {\bibfnamefont {B.}~\bibnamefont {Li}},\ }\bibfield  {title}
  {\bibinfo {title} {Topological magnon insulator in insulating ferromagnet},\
  }\href {https://doi.org/10.1103/PhysRevB.87.144101} {\bibfield  {journal}
  {\bibinfo  {journal} {Phys. Rev. B}\ }\textbf {\bibinfo {volume} {87}},\
  \bibinfo {pages} {144101} (\bibinfo {year} {2013})}\BibitemShut {NoStop}%
\bibitem [{\citenamefont {Mook}\ \emph {et~al.}(2014)\citenamefont {Mook},
  \citenamefont {Henk},\ and\ \citenamefont {Mertig}}]{Mook2014}%
  \BibitemOpen
  \bibfield  {author} {\bibinfo {author} {\bibfnamefont {A.}~\bibnamefont
  {Mook}}, \bibinfo {author} {\bibfnamefont {J.}~\bibnamefont {Henk}},\ and\
  \bibinfo {author} {\bibfnamefont {I.}~\bibnamefont {Mertig}},\ }\bibfield
  {title} {\bibinfo {title} {Edge states in topological magnon insulators},\
  }\href {https://doi.org/10.1103/PhysRevB.90.024412} {\bibfield  {journal}
  {\bibinfo  {journal} {Phys. Rev. B}\ }\textbf {\bibinfo {volume} {90}},\
  \bibinfo {pages} {024412} (\bibinfo {year} {2014})}\BibitemShut {NoStop}%
\bibitem [{\citenamefont {Owerre}(2016)}]{Owerre2016}%
  \BibitemOpen
  \bibfield  {author} {\bibinfo {author} {\bibfnamefont {S.~A.}\ \bibnamefont
  {Owerre}},\ }\bibfield  {title} {\bibinfo {title} {A first theoretical
  realization of honeycomb topological magnon insulator},\ }\href
  {https://doi.org/10.1088/0953-8984/28/38/386001} {\bibfield  {journal}
  {\bibinfo  {journal} {J. Phys.: Condens. Matter}\ }\textbf {\bibinfo {volume}
  {28}},\ \bibinfo {pages} {386001} (\bibinfo {year} {2016})}\BibitemShut
  {NoStop}%
\bibitem [{\citenamefont {Kim}\ \emph {et~al.}(2016)\citenamefont {Kim},
  \citenamefont {Ochoa}, \citenamefont {Zarzuela},\ and\ \citenamefont
  {Tserkovnyak}}]{Kim2016}%
  \BibitemOpen
  \bibfield  {author} {\bibinfo {author} {\bibfnamefont {S.~K.}\ \bibnamefont
  {Kim}}, \bibinfo {author} {\bibfnamefont {H.}~\bibnamefont {Ochoa}}, \bibinfo
  {author} {\bibfnamefont {R.}~\bibnamefont {Zarzuela}},\ and\ \bibinfo
  {author} {\bibfnamefont {Y.}~\bibnamefont {Tserkovnyak}},\ }\bibfield
  {title} {\bibinfo {title} {Realization of the haldane-kane-mele model in a
  system of localized spins},\ }\href
  {https://doi.org/10.1103/PhysRevLett.117.227201} {\bibfield  {journal}
  {\bibinfo  {journal} {Phys. Rev. Lett.}\ }\textbf {\bibinfo {volume} {117}},\
  \bibinfo {pages} {227201} (\bibinfo {year} {2016})}\BibitemShut {NoStop}%
\bibitem [{\citenamefont {Hirosawa}\ \emph {et~al.}(2020)\citenamefont
  {Hirosawa}, \citenamefont {D\'{\i}az}, \citenamefont {Klinovaja},\ and\
  \citenamefont {Loss}}]{Hirosawa2020}%
  \BibitemOpen
  \bibfield  {author} {\bibinfo {author} {\bibfnamefont {T.}~\bibnamefont
  {Hirosawa}}, \bibinfo {author} {\bibfnamefont {S.~A.}\ \bibnamefont
  {D\'{\i}az}}, \bibinfo {author} {\bibfnamefont {J.}~\bibnamefont
  {Klinovaja}},\ and\ \bibinfo {author} {\bibfnamefont {D.}~\bibnamefont
  {Loss}},\ }\bibfield  {title} {\bibinfo {title} {Magnonic quadrupole
  topological insulator in antiskyrmion crystals},\ }\href
  {https://doi.org/10.1103/PhysRevLett.125.207204} {\bibfield  {journal}
  {\bibinfo  {journal} {Phys. Rev. Lett.}\ }\textbf {\bibinfo {volume} {125}},\
  \bibinfo {pages} {207204} (\bibinfo {year} {2020})}\BibitemShut {NoStop}%
\bibitem [{\citenamefont {Shindou}\ \emph {et~al.}(2013)\citenamefont
  {Shindou}, \citenamefont {Matsumoto}, \citenamefont {Murakami},\ and\
  \citenamefont {Ohe}}]{Shindou2013}%
  \BibitemOpen
  \bibfield  {author} {\bibinfo {author} {\bibfnamefont {R.}~\bibnamefont
  {Shindou}}, \bibinfo {author} {\bibfnamefont {R.}~\bibnamefont {Matsumoto}},
  \bibinfo {author} {\bibfnamefont {S.}~\bibnamefont {Murakami}},\ and\
  \bibinfo {author} {\bibfnamefont {J.-i.}\ \bibnamefont {Ohe}},\ }\bibfield
  {title} {\bibinfo {title} {Topological chiral magnonic edge mode in a
  magnonic crystal},\ }\href {https://doi.org/10.1103/PhysRevB.87.174427}
  {\bibfield  {journal} {\bibinfo  {journal} {Phys. Rev. B}\ }\textbf {\bibinfo
  {volume} {87}},\ \bibinfo {pages} {174427} (\bibinfo {year}
  {2013})}\BibitemShut {NoStop}%
\bibitem [{\citenamefont {Li}\ \emph {et~al.}(2018)\citenamefont {Li},
  \citenamefont {Xiao},\ and\ \citenamefont {Chang}}]{Li2018}%
  \BibitemOpen
  \bibfield  {author} {\bibinfo {author} {\bibfnamefont {Y.-M.}\ \bibnamefont
  {Li}}, \bibinfo {author} {\bibfnamefont {J.}~\bibnamefont {Xiao}},\ and\
  \bibinfo {author} {\bibfnamefont {K.}~\bibnamefont {Chang}},\ }\bibfield
  {title} {\bibinfo {title} {Topological magnon modes in patterned
  ferrimagnetic insulator thin films},\ }\href
  {https://doi.org/10.1021/acs.nanolett.8b00492} {\bibfield  {journal}
  {\bibinfo  {journal} {Nano Lett.}\ }\textbf {\bibinfo {volume} {18}},\
  \bibinfo {pages} {3032} (\bibinfo {year} {2018})}\BibitemShut {NoStop}%
\bibitem [{\citenamefont {Wang}\ \emph {et~al.}(2020)\citenamefont {Wang},
  \citenamefont {Nie}, \citenamefont {Xia},\ and\ \citenamefont
  {Guo}}]{Wang2020}%
  \BibitemOpen
  \bibfield  {author} {\bibinfo {author} {\bibfnamefont {X.-G.}\ \bibnamefont
  {Wang}}, \bibinfo {author} {\bibfnamefont {Y.-Z.}\ \bibnamefont {Nie}},
  \bibinfo {author} {\bibfnamefont {Q.-L.}\ \bibnamefont {Xia}},\ and\ \bibinfo
  {author} {\bibfnamefont {G.-H.}\ \bibnamefont {Guo}},\ }\bibfield  {title}
  {\bibinfo {title} {Dynamically reconfigurable magnonic crystal composed of
  artificial magnetic skyrmion lattice},\ }\href
  {https://doi.org/10.1063/5.0012791} {\bibfield  {journal} {\bibinfo
  {journal} {J. Appl. Phys.}\ }\textbf {\bibinfo {volume} {128}},\ \bibinfo
  {pages} {063901} (\bibinfo {year} {2020})}\BibitemShut {NoStop}%
\bibitem [{\citenamefont {Klingler}\ \emph {et~al.}(2018)\citenamefont
  {Klingler}, \citenamefont {Amin}, \citenamefont {Gepr\"ags}, \citenamefont
  {Ganzhorn}, \citenamefont {Maier-Flaig}, \citenamefont {Althammer},
  \citenamefont {Huebl}, \citenamefont {Gross}, \citenamefont {McMichael},
  \citenamefont {Stiles}, \citenamefont {Goennenwein},\ and\ \citenamefont
  {Weiler}}]{Klingler2018}%
  \BibitemOpen
  \bibfield  {author} {\bibinfo {author} {\bibfnamefont {S.}~\bibnamefont
  {Klingler}}, \bibinfo {author} {\bibfnamefont {V.}~\bibnamefont {Amin}},
  \bibinfo {author} {\bibfnamefont {S.}~\bibnamefont {Gepr\"ags}}, \bibinfo
  {author} {\bibfnamefont {K.}~\bibnamefont {Ganzhorn}}, \bibinfo {author}
  {\bibfnamefont {H.}~\bibnamefont {Maier-Flaig}}, \bibinfo {author}
  {\bibfnamefont {M.}~\bibnamefont {Althammer}}, \bibinfo {author}
  {\bibfnamefont {H.}~\bibnamefont {Huebl}}, \bibinfo {author} {\bibfnamefont
  {R.}~\bibnamefont {Gross}}, \bibinfo {author} {\bibfnamefont {R.~D.}\
  \bibnamefont {McMichael}}, \bibinfo {author} {\bibfnamefont {M.~D.}\
  \bibnamefont {Stiles}}, \bibinfo {author} {\bibfnamefont {S.~T.~B.}\
  \bibnamefont {Goennenwein}},\ and\ \bibinfo {author} {\bibfnamefont
  {M.}~\bibnamefont {Weiler}},\ }\bibfield  {title} {\bibinfo {title}
  {Spin-torque excitation of perpendicular standing spin waves in coupled
  $\mathrm{YIG}/\mathrm{Co}$ heterostructures},\ }\href
  {https://doi.org/10.1103/PhysRevLett.120.127201} {\bibfield  {journal}
  {\bibinfo  {journal} {Phys. Rev. Lett.}\ }\textbf {\bibinfo {volume} {120}},\
  \bibinfo {pages} {127201} (\bibinfo {year} {2018})}\BibitemShut {NoStop}%
\bibitem [{\citenamefont {Chen}\ \emph {et~al.}(2018)\citenamefont {Chen},
  \citenamefont {Liu}, \citenamefont {Liu}, \citenamefont {Xiao}, \citenamefont
  {Xia}, \citenamefont {Bauer}, \citenamefont {Wu},\ and\ \citenamefont
  {Yu}}]{Chen2018}%
  \BibitemOpen
  \bibfield  {author} {\bibinfo {author} {\bibfnamefont {J.}~\bibnamefont
  {Chen}}, \bibinfo {author} {\bibfnamefont {C.}~\bibnamefont {Liu}}, \bibinfo
  {author} {\bibfnamefont {T.}~\bibnamefont {Liu}}, \bibinfo {author}
  {\bibfnamefont {Y.}~\bibnamefont {Xiao}}, \bibinfo {author} {\bibfnamefont
  {K.}~\bibnamefont {Xia}}, \bibinfo {author} {\bibfnamefont {G.~E.~W.}\
  \bibnamefont {Bauer}}, \bibinfo {author} {\bibfnamefont {M.}~\bibnamefont
  {Wu}},\ and\ \bibinfo {author} {\bibfnamefont {H.}~\bibnamefont {Yu}},\
  }\bibfield  {title} {\bibinfo {title} {Strong interlayer magnon-magnon
  coupling in magnetic metal-insulator hybrid nanostructures},\ }\href
  {https://doi.org/10.1103/PhysRevLett.120.217202} {\bibfield  {journal}
  {\bibinfo  {journal} {Phys. Rev. Lett.}\ }\textbf {\bibinfo {volume} {120}},\
  \bibinfo {pages} {217202} (\bibinfo {year} {2018})}\BibitemShut {NoStop}%
\bibitem [{\citenamefont {Fan}\ \emph {et~al.}(2020)\citenamefont {Fan},
  \citenamefont {Quarterman}, \citenamefont {Finley}, \citenamefont {Han},
  \citenamefont {Zhang}, \citenamefont {Hou}, \citenamefont {Stiles},
  \citenamefont {Grutter},\ and\ \citenamefont {Liu}}]{Fan2020}%
  \BibitemOpen
  \bibfield  {author} {\bibinfo {author} {\bibfnamefont {Y.}~\bibnamefont
  {Fan}}, \bibinfo {author} {\bibfnamefont {P.}~\bibnamefont {Quarterman}},
  \bibinfo {author} {\bibfnamefont {J.}~\bibnamefont {Finley}}, \bibinfo
  {author} {\bibfnamefont {J.}~\bibnamefont {Han}}, \bibinfo {author}
  {\bibfnamefont {P.}~\bibnamefont {Zhang}}, \bibinfo {author} {\bibfnamefont
  {J.~T.}\ \bibnamefont {Hou}}, \bibinfo {author} {\bibfnamefont {M.~D.}\
  \bibnamefont {Stiles}}, \bibinfo {author} {\bibfnamefont {A.~J.}\
  \bibnamefont {Grutter}},\ and\ \bibinfo {author} {\bibfnamefont
  {L.}~\bibnamefont {Liu}},\ }\bibfield  {title} {\bibinfo {title}
  {Manipulation of coupling and magnon transport in magnetic metal-insulator
  hybrid structures},\ }\href
  {https://doi.org/10.1103/PhysRevApplied.13.061002} {\bibfield  {journal}
  {\bibinfo  {journal} {Phys. Rev. Applied}\ }\textbf {\bibinfo {volume}
  {13}},\ \bibinfo {pages} {061002} (\bibinfo {year} {2020})}\BibitemShut
  {NoStop}%
\bibitem [{SM()}]{SM}%
  \BibitemOpen
  \href@noop {} {}\bibinfo {note} {See Supplemental Material for more
  details}\BibitemShut {NoStop}%
\bibitem [{\citenamefont {Yu}\ \emph {et~al.}(2019)\citenamefont {Yu},
  \citenamefont {Liu}, \citenamefont {Yu}, \citenamefont {Blanter},\ and\
  \citenamefont {Bauer}}]{Yu2019}%
  \BibitemOpen
  \bibfield  {author} {\bibinfo {author} {\bibfnamefont {T.}~\bibnamefont
  {Yu}}, \bibinfo {author} {\bibfnamefont {C.}~\bibnamefont {Liu}}, \bibinfo
  {author} {\bibfnamefont {H.}~\bibnamefont {Yu}}, \bibinfo {author}
  {\bibfnamefont {Y.~M.}\ \bibnamefont {Blanter}},\ and\ \bibinfo {author}
  {\bibfnamefont {G.~E.~W.}\ \bibnamefont {Bauer}},\ }\bibfield  {title}
  {\bibinfo {title} {Chiral excitation of spin waves in ferromagnetic films by
  magnetic nanowire gratings},\ }\href
  {https://doi.org/10.1103/PhysRevB.99.134424} {\bibfield  {journal} {\bibinfo
  {journal} {Phys. Rev. B}\ }\textbf {\bibinfo {volume} {99}},\ \bibinfo
  {pages} {134424} (\bibinfo {year} {2019})}\BibitemShut {NoStop}%
\bibitem [{\citenamefont {Chen}\ \emph {et~al.}(2019)\citenamefont {Chen},
  \citenamefont {Yu}, \citenamefont {Liu}, \citenamefont {Liu}, \citenamefont
  {Madami}, \citenamefont {Shen}, \citenamefont {Zhang}, \citenamefont {Tu},
  \citenamefont {Alam}, \citenamefont {Xia}, \citenamefont {Wu}, \citenamefont
  {Gubbiotti}, \citenamefont {Blanter}, \citenamefont {Bauer},\ and\
  \citenamefont {Yu}}]{Chen2019}%
  \BibitemOpen
  \bibfield  {author} {\bibinfo {author} {\bibfnamefont {J.}~\bibnamefont
  {Chen}}, \bibinfo {author} {\bibfnamefont {T.}~\bibnamefont {Yu}}, \bibinfo
  {author} {\bibfnamefont {C.}~\bibnamefont {Liu}}, \bibinfo {author}
  {\bibfnamefont {T.}~\bibnamefont {Liu}}, \bibinfo {author} {\bibfnamefont
  {M.}~\bibnamefont {Madami}}, \bibinfo {author} {\bibfnamefont
  {K.}~\bibnamefont {Shen}}, \bibinfo {author} {\bibfnamefont {J.}~\bibnamefont
  {Zhang}}, \bibinfo {author} {\bibfnamefont {S.}~\bibnamefont {Tu}}, \bibinfo
  {author} {\bibfnamefont {M.~S.}\ \bibnamefont {Alam}}, \bibinfo {author}
  {\bibfnamefont {K.}~\bibnamefont {Xia}}, \bibinfo {author} {\bibfnamefont
  {M.}~\bibnamefont {Wu}}, \bibinfo {author} {\bibfnamefont {G.}~\bibnamefont
  {Gubbiotti}}, \bibinfo {author} {\bibfnamefont {Y.~M.}\ \bibnamefont
  {Blanter}}, \bibinfo {author} {\bibfnamefont {G.~E.~W.}\ \bibnamefont
  {Bauer}},\ and\ \bibinfo {author} {\bibfnamefont {H.}~\bibnamefont {Yu}},\
  }\bibfield  {title} {\bibinfo {title} {Excitation of unidirectional exchange
  spin waves by a nanoscale magnetic grating},\ }\href
  {https://doi.org/10.1103/PhysRevB.100.104427} {\bibfield  {journal} {\bibinfo
   {journal} {Phys. Rev. B}\ }\textbf {\bibinfo {volume} {100}},\ \bibinfo
  {pages} {104427} (\bibinfo {year} {2019})}\BibitemShut {NoStop}%
\bibitem [{\citenamefont {Ishibashi}\ \emph {et~al.}(2020)\citenamefont
  {Ishibashi}, \citenamefont {Shiota}, \citenamefont {Li}, \citenamefont
  {Funada}, \citenamefont {Moriyama},\ and\ \citenamefont
  {Ono}}]{Ishibashi2020}%
  \BibitemOpen
  \bibfield  {author} {\bibinfo {author} {\bibfnamefont {M.}~\bibnamefont
  {Ishibashi}}, \bibinfo {author} {\bibfnamefont {Y.}~\bibnamefont {Shiota}},
  \bibinfo {author} {\bibfnamefont {T.}~\bibnamefont {Li}}, \bibinfo {author}
  {\bibfnamefont {S.}~\bibnamefont {Funada}}, \bibinfo {author} {\bibfnamefont
  {T.}~\bibnamefont {Moriyama}},\ and\ \bibinfo {author} {\bibfnamefont
  {T.}~\bibnamefont {Ono}},\ }\bibfield  {title} {\bibinfo {title} {Switchable
  giant nonreciprocal frequency shift of propagating spin waves in synthetic
  antiferromagnets},\ }\href {https://doi.org/10.1126/sciadv.aaz6931}
  {\bibfield  {journal} {\bibinfo  {journal} {Sci. Adv.}\ }\textbf {\bibinfo
  {volume} {6}},\ \bibinfo {pages} {eaaz6931} (\bibinfo {year}
  {2020})}\BibitemShut {NoStop}%
\bibitem [{\citenamefont {Han}\ \emph {et~al.}(2021)\citenamefont {Han},
  \citenamefont {Fan}, \citenamefont {McGoldrick}, \citenamefont {Finley},
  \citenamefont {Hou}, \citenamefont {Zhang},\ and\ \citenamefont
  {Liu}}]{Han2021}%
  \BibitemOpen
  \bibfield  {author} {\bibinfo {author} {\bibfnamefont {J.}~\bibnamefont
  {Han}}, \bibinfo {author} {\bibfnamefont {Y.}~\bibnamefont {Fan}}, \bibinfo
  {author} {\bibfnamefont {B.~C.}\ \bibnamefont {McGoldrick}}, \bibinfo
  {author} {\bibfnamefont {J.}~\bibnamefont {Finley}}, \bibinfo {author}
  {\bibfnamefont {J.~T.}\ \bibnamefont {Hou}}, \bibinfo {author} {\bibfnamefont
  {P.}~\bibnamefont {Zhang}},\ and\ \bibinfo {author} {\bibfnamefont
  {L.}~\bibnamefont {Liu}},\ }\bibfield  {title} {\bibinfo {title}
  {Nonreciprocal transmission of incoherent magnons with asymmetric diffusion
  length},\ }\href {https://doi.org/10.1021/acs.nanolett.1c02575} {\bibfield
  {journal} {\bibinfo  {journal} {Nano Lett.}\ }\textbf {\bibinfo {volume}
  {21}},\ \bibinfo {pages} {7037} (\bibinfo {year} {2021})}\BibitemShut
  {NoStop}%
\bibitem [{\citenamefont {Su}\ \emph {et~al.}(1979)\citenamefont {Su},
  \citenamefont {Schrieffer},\ and\ \citenamefont {Heeger}}]{Su1979}%
  \BibitemOpen
  \bibfield  {author} {\bibinfo {author} {\bibfnamefont {W.~P.}\ \bibnamefont
  {Su}}, \bibinfo {author} {\bibfnamefont {J.~R.}\ \bibnamefont {Schrieffer}},\
  and\ \bibinfo {author} {\bibfnamefont {A.~J.}\ \bibnamefont {Heeger}},\
  }\bibfield  {title} {\bibinfo {title} {Solitons in polyacetylene},\ }\href
  {https://doi.org/10.1103/PhysRevLett.42.1698} {\bibfield  {journal} {\bibinfo
   {journal} {Phys. Rev. Lett.}\ }\textbf {\bibinfo {volume} {42}},\ \bibinfo
  {pages} {1698} (\bibinfo {year} {1979})}\BibitemShut {NoStop}%
\bibitem [{\citenamefont {Klingler}\ \emph {et~al.}(2014)\citenamefont
  {Klingler}, \citenamefont {Chumak}, \citenamefont {Mewes}, \citenamefont
  {Khodadadi}, \citenamefont {Mewes}, \citenamefont {Dubs}, \citenamefont
  {Surzhenko}, \citenamefont {Hillebrands},\ and\ \citenamefont
  {Conca}}]{Klingler2014}%
  \BibitemOpen
  \bibfield  {author} {\bibinfo {author} {\bibfnamefont {S.}~\bibnamefont
  {Klingler}}, \bibinfo {author} {\bibfnamefont {A.}~\bibnamefont {Chumak}},
  \bibinfo {author} {\bibfnamefont {T.}~\bibnamefont {Mewes}}, \bibinfo
  {author} {\bibfnamefont {B.}~\bibnamefont {Khodadadi}}, \bibinfo {author}
  {\bibfnamefont {C.}~\bibnamefont {Mewes}}, \bibinfo {author} {\bibfnamefont
  {C.}~\bibnamefont {Dubs}}, \bibinfo {author} {\bibfnamefont {O.}~\bibnamefont
  {Surzhenko}}, \bibinfo {author} {\bibfnamefont {B.}~\bibnamefont
  {Hillebrands}},\ and\ \bibinfo {author} {\bibfnamefont {A.}~\bibnamefont
  {Conca}},\ }\bibfield  {title} {\bibinfo {title} {Measurements of the
  exchange stiffness of yig films using broadband ferromagnetic resonance
  techniques},\ }\href {https://doi.org/10.1088/0022-3727/48/1/015001}
  {\bibfield  {journal} {\bibinfo  {journal} {J. Phys. D: Appl. Phys.}\
  }\textbf {\bibinfo {volume} {48}},\ \bibinfo {pages} {015001} (\bibinfo
  {year} {2014})}\BibitemShut {NoStop}%
\bibitem [{\citenamefont {Langer}\ \emph {et~al.}(2016)\citenamefont {Langer},
  \citenamefont {Wagner}, \citenamefont {Sebastian}, \citenamefont
  {H{\"u}bner}, \citenamefont {Grenzer}, \citenamefont {Wang}, \citenamefont
  {Kubota}, \citenamefont {Schneider}, \citenamefont {Stienen}, \citenamefont
  {Lenz}, \citenamefont {Schulthei\ss}, \citenamefont {Lindner}, \citenamefont
  {Takanashi}, \citenamefont {Arias},\ and\ \citenamefont
  {Fassbender}}]{Langer2016}%
  \BibitemOpen
  \bibfield  {author} {\bibinfo {author} {\bibfnamefont {M.}~\bibnamefont
  {Langer}}, \bibinfo {author} {\bibfnamefont {K.}~\bibnamefont {Wagner}},
  \bibinfo {author} {\bibfnamefont {T.}~\bibnamefont {Sebastian}}, \bibinfo
  {author} {\bibfnamefont {R.}~\bibnamefont {H{\"u}bner}}, \bibinfo {author}
  {\bibfnamefont {J.}~\bibnamefont {Grenzer}}, \bibinfo {author} {\bibfnamefont
  {Y.}~\bibnamefont {Wang}}, \bibinfo {author} {\bibfnamefont {T.}~\bibnamefont
  {Kubota}}, \bibinfo {author} {\bibfnamefont {T.}~\bibnamefont {Schneider}},
  \bibinfo {author} {\bibfnamefont {S.}~\bibnamefont {Stienen}}, \bibinfo
  {author} {\bibfnamefont {K.}~\bibnamefont {Lenz}}, \bibinfo {author}
  {\bibfnamefont {H.}~\bibnamefont {Schulthei\ss}}, \bibinfo {author}
  {\bibfnamefont {J.}~\bibnamefont {Lindner}}, \bibinfo {author} {\bibfnamefont
  {K.}~\bibnamefont {Takanashi}}, \bibinfo {author} {\bibfnamefont {R.~E.}\
  \bibnamefont {Arias}},\ and\ \bibinfo {author} {\bibfnamefont
  {J.}~\bibnamefont {Fassbender}},\ }\bibfield  {title} {\bibinfo {title}
  {Parameter-free determination of the exchange constant in thin films using
  magnonic patterning},\ }\href {https://doi.org/10.1063/1.4943228} {\bibfield
  {journal} {\bibinfo  {journal} {Appl. Phys. Lett.}\ }\textbf {\bibinfo
  {volume} {108}},\ \bibinfo {pages} {102402} (\bibinfo {year}
  {2016})}\BibitemShut {NoStop}%
\bibitem [{\citenamefont {Kane}\ \emph {et~al.}(2002)\citenamefont {Kane},
  \citenamefont {Mukhopadhyay},\ and\ \citenamefont {Lubensky}}]{Kane2002}%
  \BibitemOpen
  \bibfield  {author} {\bibinfo {author} {\bibfnamefont {C.~L.}\ \bibnamefont
  {Kane}}, \bibinfo {author} {\bibfnamefont {R.}~\bibnamefont {Mukhopadhyay}},\
  and\ \bibinfo {author} {\bibfnamefont {T.~C.}\ \bibnamefont {Lubensky}},\
  }\bibfield  {title} {\bibinfo {title} {Fractional quantum hall effect in an
  array of quantum wires},\ }\href
  {https://doi.org/10.1103/PhysRevLett.88.036401} {\bibfield  {journal}
  {\bibinfo  {journal} {Phys. Rev. Lett.}\ }\textbf {\bibinfo {volume} {88}},\
  \bibinfo {pages} {036401} (\bibinfo {year} {2002})}\BibitemShut {NoStop}%
\bibitem [{\citenamefont {Meng}\ \emph {et~al.}(2015)\citenamefont {Meng},
  \citenamefont {Neupert}, \citenamefont {Greiter},\ and\ \citenamefont
  {Thomale}}]{Meng2015}%
  \BibitemOpen
  \bibfield  {author} {\bibinfo {author} {\bibfnamefont {T.}~\bibnamefont
  {Meng}}, \bibinfo {author} {\bibfnamefont {T.}~\bibnamefont {Neupert}},
  \bibinfo {author} {\bibfnamefont {M.}~\bibnamefont {Greiter}},\ and\ \bibinfo
  {author} {\bibfnamefont {R.}~\bibnamefont {Thomale}},\ }\bibfield  {title}
  {\bibinfo {title} {Coupled-wire construction of chiral spin liquids},\ }\href
  {https://doi.org/10.1103/PhysRevB.91.241106} {\bibfield  {journal} {\bibinfo
  {journal} {Phys. Rev. B}\ }\textbf {\bibinfo {volume} {91}},\ \bibinfo
  {pages} {241106} (\bibinfo {year} {2015})}\BibitemShut {NoStop}%
\bibitem [{\citenamefont {Kane}\ and\ \citenamefont {Stern}(2018)}]{Kane2018}%
  \BibitemOpen
  \bibfield  {author} {\bibinfo {author} {\bibfnamefont {C.~L.}\ \bibnamefont
  {Kane}}\ and\ \bibinfo {author} {\bibfnamefont {A.}~\bibnamefont {Stern}},\
  }\bibfield  {title} {\bibinfo {title} {Coupled wire model of ${Z}_{4}$
  orbifold quantum hall states},\ }\href
  {https://doi.org/10.1103/PhysRevB.98.085302} {\bibfield  {journal} {\bibinfo
  {journal} {Phys. Rev. B}\ }\textbf {\bibinfo {volume} {98}},\ \bibinfo
  {pages} {085302} (\bibinfo {year} {2018})}\BibitemShut {NoStop}%
\bibitem [{\citenamefont {Wu}\ \emph {et~al.}(2019)\citenamefont {Wu},
  \citenamefont {Jian},\ and\ \citenamefont {Xu}}]{Wu2019}%
  \BibitemOpen
  \bibfield  {author} {\bibinfo {author} {\bibfnamefont {X.-C.}\ \bibnamefont
  {Wu}}, \bibinfo {author} {\bibfnamefont {C.-M.}\ \bibnamefont {Jian}},\ and\
  \bibinfo {author} {\bibfnamefont {C.}~\bibnamefont {Xu}},\ }\bibfield
  {title} {\bibinfo {title} {Coupled-wire description of the correlated physics
  in twisted bilayer graphene},\ }\href
  {https://doi.org/10.1103/PhysRevB.99.161405} {\bibfield  {journal} {\bibinfo
  {journal} {Phys. Rev. B}\ }\textbf {\bibinfo {volume} {99}},\ \bibinfo
  {pages} {161405} (\bibinfo {year} {2019})}\BibitemShut {NoStop}%
\bibitem [{\citenamefont {Colpa}(1978)}]{Colpa1978}%
  \BibitemOpen
  \bibfield  {author} {\bibinfo {author} {\bibfnamefont {J.~H.~P.}\
  \bibnamefont {Colpa}},\ }\bibfield  {title} {\bibinfo {title}
  {Diagonalization of the quadratic boson hamiltonian},\ }\href
  {https://doi.org/10.1016/0378-4371(78)90160-7} {\bibfield  {journal}
  {\bibinfo  {journal} {Phys. A: Stat. Mech. Appl.}\ }\textbf {\bibinfo
  {volume} {93}},\ \bibinfo {pages} {327} (\bibinfo {year} {1978})}\BibitemShut
  {NoStop}%
\bibitem [{\citenamefont {Fu}\ \emph {et~al.}(2007)\citenamefont {Fu},
  \citenamefont {Kane},\ and\ \citenamefont {Mele}}]{Fu2007}%
  \BibitemOpen
  \bibfield  {author} {\bibinfo {author} {\bibfnamefont {L.}~\bibnamefont
  {Fu}}, \bibinfo {author} {\bibfnamefont {C.~L.}\ \bibnamefont {Kane}},\ and\
  \bibinfo {author} {\bibfnamefont {E.~J.}\ \bibnamefont {Mele}},\ }\bibfield
  {title} {\bibinfo {title} {Topological insulators in three dimensions},\
  }\href {https://doi.org/10.1103/PhysRevLett.98.106803} {\bibfield  {journal}
  {\bibinfo  {journal} {Phys. Rev. Lett.}\ }\textbf {\bibinfo {volume} {98}},\
  \bibinfo {pages} {106803} (\bibinfo {year} {2007})}\BibitemShut {NoStop}%
\end{thebibliography}%


\providecommand{\noopsort}[1]{}\providecommand{\singleletter}[1]{#1}%
\begin{thebibliography}{4}%
\makeatletter
\providecommand \@ifxundefined [1]{%
 \@ifx{#1\undefined}
}%
\providecommand \@ifnum [1]{%
 \ifnum #1\expandafter \@firstoftwo
 \else \expandafter \@secondoftwo
 \fi
}%
\providecommand \@ifx [1]{%
 \ifx #1\expandafter \@firstoftwo
 \else \expandafter \@secondoftwo
 \fi
}%
\providecommand \natexlab [1]{#1}%
\providecommand \enquote  [1]{``#1''}%
\providecommand \bibnamefont  [1]{#1}%
\providecommand \bibfnamefont [1]{#1}%
\providecommand \citenamefont [1]{#1}%
\providecommand \href@noop [0]{\@secondoftwo}%
\providecommand \href [0]{\begingroup \@sanitize@url \@href}%
\providecommand \@href[1]{\@@startlink{#1}\@@href}%
\providecommand \@@href[1]{\endgroup#1\@@endlink}%
\providecommand \@sanitize@url [0]{\catcode `\\12\catcode `\$12\catcode
  `\&12\catcode `\#12\catcode `\^12\catcode `\_12\catcode `\%12\relax}%
\providecommand \@@startlink[1]{}%
\providecommand \@@endlink[0]{}%
\providecommand \url  [0]{\begingroup\@sanitize@url \@url }%
\providecommand \@url [1]{\endgroup\@href {#1}{\urlprefix }}%
\providecommand \urlprefix  [0]{URL }%
\providecommand \Eprint [0]{\href }%
\providecommand \doibase [0]{https://doi.org/}%
\providecommand \selectlanguage [0]{\@gobble}%
\providecommand \bibinfo  [0]{\@secondoftwo}%
\providecommand \bibfield  [0]{\@secondoftwo}%
\providecommand \translation [1]{[#1]}%
\providecommand \BibitemOpen [0]{}%
\providecommand \bibitemStop [0]{}%
\providecommand \bibitemNoStop [0]{.\EOS\space}%
\providecommand \EOS [0]{\spacefactor3000\relax}%
\providecommand \BibitemShut  [1]{\csname bibitem#1\endcsname}%
\let\auto@bib@innerbib\@empty
\bibitem [{\citenamefont {Cohen}\ and\ \citenamefont
  {Keffer}(1955)}]{Cohen1955}%
  \BibitemOpen
  \bibfield  {author} {\bibinfo {author} {\bibfnamefont {M.~H.}\ \bibnamefont
  {Cohen}}\ and\ \bibinfo {author} {\bibfnamefont {F.}~\bibnamefont {Keffer}},\
  }\href {https://doi.org/10.1103/PhysRev.99.1128} {\bibfield  {journal}
  {\bibinfo  {journal} {Phys. Rev.}\ }\textbf {\bibinfo {volume} {99}},\
  \bibinfo {pages} {1128} (\bibinfo {year} {1955})}\BibitemShut {NoStop}%
\bibitem [{\citenamefont {Berkowitz}\ and\ \citenamefont
  {Takano}(1999)}]{Berkowitz1999}%
  \BibitemOpen
  \bibfield  {author} {\bibinfo {author} {\bibfnamefont {A.}~\bibnamefont
  {Berkowitz}}\ and\ \bibinfo {author} {\bibfnamefont {K.}~\bibnamefont
  {Takano}},\ }\href
  {https://doi.org/https://doi.org/10.1016/S0304-8853(99)00453-9} {\bibfield
  {journal} {\bibinfo  {journal} {J. Magn. Magn. Mater.}\ }\textbf {\bibinfo
  {volume} {200}},\ \bibinfo {pages} {552} (\bibinfo {year}
  {1999})}\BibitemShut {NoStop}%
\bibitem [{\citenamefont {Klingler}\ \emph {et~al.}(2014)\citenamefont
  {Klingler}, \citenamefont {Chumak}, \citenamefont {Mewes}, \citenamefont
  {Khodadadi}, \citenamefont {Mewes}, \citenamefont {Dubs}, \citenamefont
  {Surzhenko}, \citenamefont {Hillebrands},\ and\ \citenamefont
  {Conca}}]{Klingler2014}%
  \BibitemOpen
  \bibfield  {author} {\bibinfo {author} {\bibfnamefont {S.}~\bibnamefont
  {Klingler}}, \bibinfo {author} {\bibfnamefont {A.~V.}\ \bibnamefont
  {Chumak}}, \bibinfo {author} {\bibfnamefont {T.}~\bibnamefont {Mewes}},
  \bibinfo {author} {\bibfnamefont {B.}~\bibnamefont {Khodadadi}}, \bibinfo
  {author} {\bibfnamefont {C.}~\bibnamefont {Mewes}}, \bibinfo {author}
  {\bibfnamefont {C.}~\bibnamefont {Dubs}}, \bibinfo {author} {\bibfnamefont
  {O.}~\bibnamefont {Surzhenko}}, \bibinfo {author} {\bibfnamefont
  {B.}~\bibnamefont {Hillebrands}},\ and\ \bibinfo {author} {\bibfnamefont
  {A.}~\bibnamefont {Conca}},\ }\href
  {https://doi.org/10.1088/0022-3727/48/1/015001} {\bibfield  {journal}
  {\bibinfo  {journal} {J. Phys. D: Appl. Phys.}\ }\textbf {\bibinfo {volume}
  {48}},\ \bibinfo {pages} {015001} (\bibinfo {year} {2014})}\BibitemShut
  {NoStop}%
\bibitem [{\citenamefont {Holstein}\ and\ \citenamefont
  {Primakoff}(1940)}]{Holstein1940}%
  \BibitemOpen
  \bibfield  {author} {\bibinfo {author} {\bibfnamefont {T.}~\bibnamefont
  {Holstein}}\ and\ \bibinfo {author} {\bibfnamefont {H.}~\bibnamefont
  {Primakoff}},\ }\href {https://doi.org/10.1103/PhysRev.58.1098} {\bibfield
  {journal} {\bibinfo  {journal} {Phys. Rev.}\ }\textbf {\bibinfo {volume}
  {58}},\ \bibinfo {pages} {1098} (\bibinfo {year} {1940})}\BibitemShut
  {NoStop}%
\end{thebibliography}%

\end{document}


\title{Supplemental Material: Tunable Magnonic Chern Bands and Chiral Spin Currents in Magnetic Multilayers}

\author{Zhongqiang Hu}
\email{zhongqhu@mit.edu}
\affiliation{Department of Electrical Engineering and Computer Science, Massachusetts Institute of Technology, Cambridge, MA 02139, USA}

\author{Liang Fu}
\affiliation{Department of Physics, Massachusetts Institute of Technology, Cambridge, MA 02139, USA}

\author{Luqiao Liu}
\affiliation{Department of Electrical Engineering and Computer Science, Massachusetts Institute of Technology, Cambridge, MA 02139, USA}

\date{\today}

\maketitle

\appendix

\section{\label{app:A} Interlayer dipolar field from propagating magnons}

\begin{figure} [t]
	\includegraphics[width=1\textwidth]{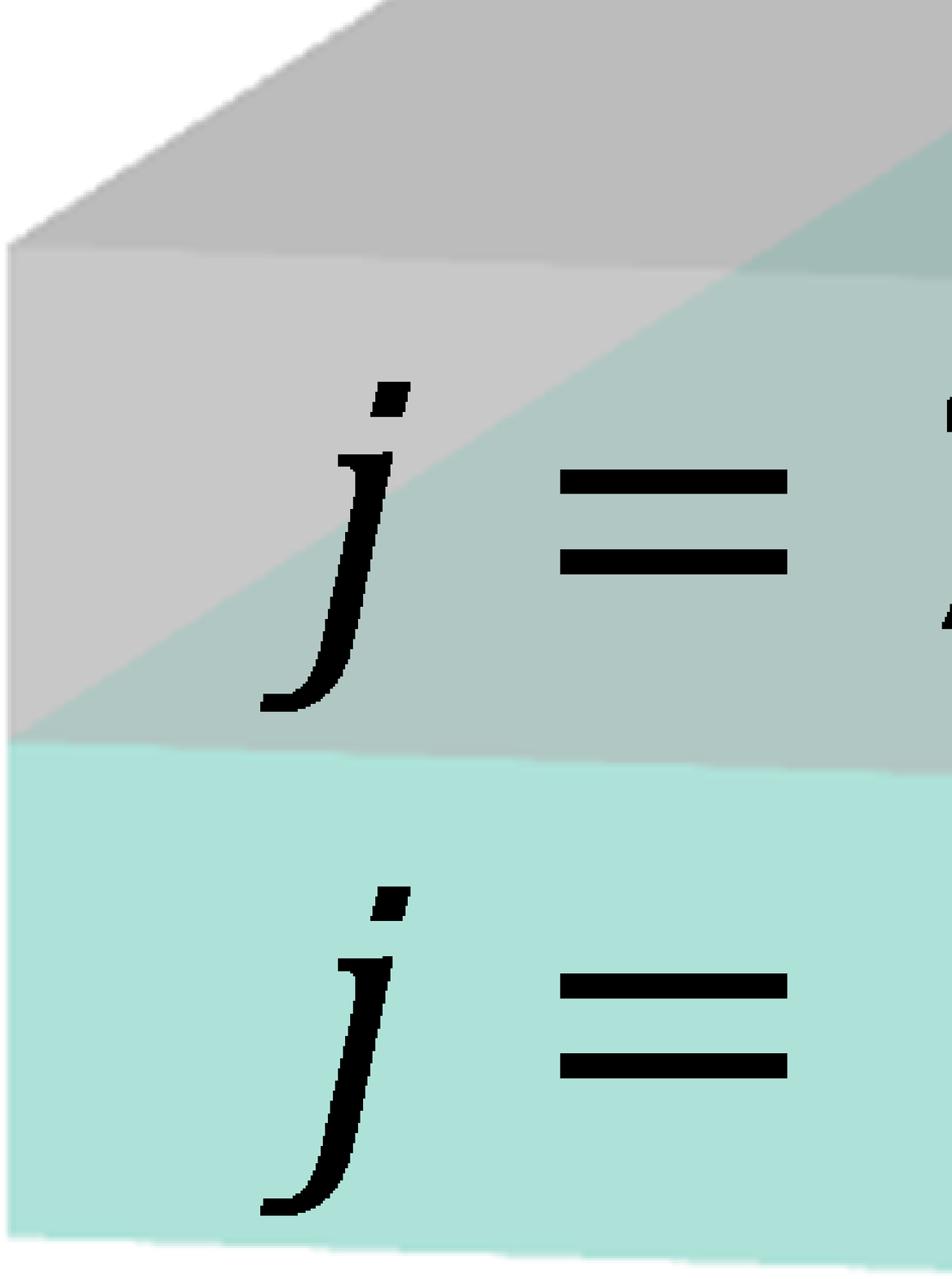}
	\caption{(a) Magnetic bilayers consisting of two different ferromagnetic layers with in-plane magnetic moments. (b) The global and local coordinate systems, as seen from the top view of the bilayers. The equilibrium moment orientations in two layers have different angles with the $x$ axis. (c) When calculating the interlayer dipolar energy, $\bm{r}$ and $\bm{r}^\prime$ are in the different layers. (d) When calculating the intralayer dipolar energy, $\bm{r}$ and $\bm{r}^\prime$ are in the same layer.}
	\label{fig:S1}
\end{figure}

We first discuss the magnetic bilayers shown in Fig. \hyperref[fig:S1]{S1(a)}, which consist of two different ferromagnetic layers with in-plane magnetic moments. For convenience, we set up a global coordinate system $(x, y, z)$, and denote the angle between the equilibrium moment orientation and the $x$ axis as $\theta_{j}$  [Fig. \hyperref[fig:S1]{S1(b)}]. Here, $j = 1, 2$ is the layer index. In the discussion throughout the paper including the main text and Supplemental Material, we use the dimensionless reduced moment $\bm{m}_{j} = \bm{M}_{j} / M_{{\rm s}j}$, where $M_{{\rm s}j}$ is the saturated magnetization. Within each layer, we also set up a local coordinate system ($x_{j}, y_{j}, z$), with the equilibrium moment orientation chosen as the $y_j$ axis.

The dipolar energy between a pair of moments $\bm{m}_1$ (located at $\bm{r}$) and $\bm{m}_2$ (located at $\bm{r}^\prime$) is
\begin{equation}
\delta E_{\rm d} = -\frac{\mu_0M_{{\rm s1}}M_{{\rm s2}}}{4\pi}\frac{3(\bm{m}_1 \cdot \bm{R})(\bm{m}_2 \cdot \bm{R}) - (\bm{m}_1 \cdot \bm{m}_2)R^2}{R^5},
\label{pair dip}
\end{equation}

\noindent where $\bm{R} = \bm{r} - \bm{r}^\prime = (X, Y, Z)$. If defining a matrix
\begin{equation}
\begin{aligned}
\bm{F} = -\frac{\mu_0M_{{\rm s1}}M_{{\rm s2}}}{4\pi R^5}
\begin{bmatrix}
3X^2 - R^2 & 3XY & 3XZ \\
3XY & 3Y^2 - R^2 & 3YZ \\
3XZ & 3YZ & 3Z^2 - R^2
\end{bmatrix}
\end{aligned},
\label{F matrix}
\end{equation}

\noindent we can write Eq. \eqref{pair dip} as $\delta E_{\rm d} = \bm{m}_2^T \bm{F} \bm{m}_1$. We note that here $\bm{m}_{j} = [m_{jx}, m_{jy}, m_{jz}]^T$ is defined as a column vector. As shown in Fig. \hyperref[fig:S1]{S1(c)}, the dipolar energy between two layers (i.e., a pair of continuous bodies) is
\begin{equation}
\begin{aligned}
E_{{\rm d}} = \iint\bm{m}_2^T(\bm{r^\prime}) \bm{F}(\bm{r - r^\prime}) \bm{m}_1(\bm{r}) d^3\bm{r}d^3\bm{r^\prime}
\end{aligned}.
\label{Ed original}
\end{equation}

\noindent After the Fourier transformation $\tilde{\bm{m}}_j = \mathcal{F}\{\bm{m}_j\}$ and $\tilde{\bm{F}} = \mathcal{F}\{\bm{F}\}$, it can be written as
\begin{equation}
\begin{aligned}
E_{{\rm d}} = \int \tilde{\bm{m}}_2^{\dagger}(\bm{k}) \tilde{\bm{F}}(\bm{k}) \tilde{\bm{m}}_1(\bm{k}) \frac{d^3\bm{k}}{(2\pi)^3}
\end{aligned}.
\label{Ed kspace}
\end{equation}

\noindent According to \cite{Cohen1955}, $\tilde{\bm{F}}(\bm{k})$ is approximated as
\begin{equation}
\begin{aligned}
\tilde{\bm{F}}(\bm{k}) \approx -\frac{\mu_0M_{{\rm s1}}M_{{\rm s2}}}{3k^2}
\begin{bmatrix}
3k_x^2 - k^2 & 3k_xk_y & 3k_xk_z \\
3k_xk_y & 3k_y^2 - k^2 & 3k_yk_z \\
3k_xk_z & 3k_yk_z & 3k_z^2 - k^2
\end{bmatrix}
\end{aligned},
\label{F matrix kspace}
\end{equation}

\noindent when $k = |\bm{k}|$ is not large compared with the Brillion zone boundary. Considering the thin-film configuration with the lowest standing mode along the $z$ direction, i.e., $\bm{m}_j (\bm{r}) =  \bm{m}_j(\bm{r}_\parallel) S_0(z; d_{j})$, where $d_j$ is the layer thickness and
\begin{equation}
\begin{gathered}
S_0(z; d_{1}) = \left\{\begin{aligned}
	1 \quad &(-d_1 < z < 0) \\
	0 \quad &({\rm otherwise}),
	\end{aligned} \right. \\
S_0(z; d_{2}) = \left\{\begin{aligned}
	1 \quad &(0 < z < d_2) \\
	0 \quad &({\rm otherwise}),
	\end{aligned} \right.
\end{gathered}
\label{S0}
\end{equation}

\noindent we can rewrite Eq. \eqref{Ed kspace} as
\begin{equation}
\begin{aligned}
E_{{\rm d}} = \int \tilde{\bm{m}}_2^{\dagger}(\bm{k}_\parallel)\tilde{S}_0^*(k_z;d_2) \tilde{\bm{F}}(\bm{k}) \tilde{S}_0(k_z;d_1)\tilde{\bm{m}}_1(\bm{k}_\parallel) \frac{d^3\bm{k}}{(2\pi)^3} 
= \int \tilde{\bm{m}}_2^{\dagger}(\bm{k}_\parallel) \bm{G}(\bm{k}_\parallel) \tilde{\bm{m}}_1(\bm{k}_\parallel) \frac{d^2\bm{k}_\parallel}{(2\pi)^2}, 
\end{aligned}
\label{Ed kparallel}
\end{equation}

\noindent where
\begin{equation}
\begin{gathered}
\tilde{S}_0(k_z;d_1) = \mathcal{F}\{S_0(z; d_{1})\} = d_1 {e^{ik_zd_1/2}} {\rm sinc}(k_zd_1/2), \\
\tilde{S}_0(k_z;d_2) = \mathcal{F}\{S_0(z; d_{2})\} = d_2 e^{-ik_zd_2/2} {\rm sinc}(k_zd_2/2), \\
\end{gathered}
\label{S0 kspace}
\end{equation}

\noindent and
\begin{equation}
\begin{aligned}
\bm{G}({\bm{k}_\parallel}) = \int\tilde{S}_0^*(k_z;d_2) \tilde{\bm{F}}(\bm{k})\tilde{S}_0(k_z;d_1) \frac{dk_z}{2\pi}
= \frac{\mu_0M_{{\rm s1}}M_{{\rm s2}}d_1d_2}{2k_\parallel} N(k_\parallel;d_1) N(k_\parallel;d_2)
\begin{bmatrix}
-k_x^2 & -k_xk_y & ik_xk_\parallel \\
-k_xk_y & -k_y^2 & ik_yk_\parallel \\
ik_xk_\parallel & ik_yk_\parallel & k^2_\parallel
\end{bmatrix},
\end{aligned}
\label{G kparallel}
\end{equation}

\noindent where $N$ has a form of $N(k; d) = (1 - e^{-kd})/(kd)$. We note that in Eq. \eqref{Ed kparallel}, the moments $\tilde{\bm{m}}_1$ and $\tilde{\bm{m}}_2$ are expressed in the global coordinate system. After switching to local coordinate systems, we would get a new matrix $\bm{g}(\bm{k}_\parallel)$ instead of $\bm{G}(\bm{k}_\parallel)$ to express $E_{\rm d}$:
\begin{equation}
\begin{aligned}
E_{{\rm d}} = \int \tilde{\bm{m}}_2^{\dagger}(\bm{k}_\parallel) \bm{g}(\bm{k}_\parallel) \tilde{\bm{m}}_1(\bm{k}_\parallel) \frac{d^2\bm{k}_\parallel}{(2\pi)^2}
\end{aligned}.
\label{Ed g}
\end{equation}

\noindent Defining the rotation matrix as
\begin{equation}
R_{j} = \begin{bmatrix}
{\rm sin}\theta_j & {\rm cos}\theta_j & 0 \\
-{\rm cos}\theta_j & {\rm sin}\theta_j & 0 \\
0 & 0 & 1
\end{bmatrix},
\label{rot matrix}
\end{equation}

\noindent and only considering magnons confined to transport in the $x$ axis, which means $k_\parallel = |k_x|$ and $k_y = 0$ (i.e., the simplified case mentioned in the main text, where each layer is a infinitely long strip along the $x$ axis and the magnetic moment distribution along the $y$ axis is uniform), we get
\begin{equation}
\begin{aligned}
\bm{g}(k_x) = R^{-1}_2 \bm{G}(k_x) R_1
= \frac{1}{2}\mu_0M_{{\rm s1}}M_{{\rm s2}}d_1d_2 N(|k_x|;d_1) N(|k_x|;d_2)
\begin{bmatrix}
-|k_x|{\rm sin}\theta_1{\rm sin}\theta_2 & -|k_x|{\rm cos}\theta_1{\rm sin}\theta_2 & ik_x{\rm sin}\theta_2 \\
-|k_x|{\rm sin}\theta_1{\rm cos}\theta_2 & -|k_x|{\rm cos}\theta_1{\rm cos}\theta_2 & ik_x{\rm cos}\theta_2 \\
ik_x{\rm sin}\theta_1 & ik_x{\rm cos}\theta_1 & |k_x|
\end{bmatrix}.
\end{aligned}
\label{g kx}
\end{equation}

\noindent Right now, the interlayer dipolar energy is written as
\begin{equation}
\begin{aligned}
E_{{\rm d}} = \int \tilde{\bm{m}}_2^{\dagger}(k_x) \bm{g}(k_x) \tilde{\bm{m}}_1(k_x) \frac{dk_x}{2\pi}
= -\mu_0M_{{\rm s}1}d_1 \int \tilde{\bm{h}}_1^{{\rm d}\dagger}(k_x)\tilde{\bm{m}}_1(k_x) \frac{dk_x}{2\pi}
=-\mu_0M_{{\rm s}2}d_2 \int \tilde{\bm{m}}_2^{\dagger}(k_x)\tilde{\bm{h}}_2^{{\rm d}}(k_x) \frac{dk_x}{2\pi},
\end{aligned}
\label{Ed hd}
\end{equation}

\noindent and the interlayer dipolar field $\tilde{\bm{h}}_j^{{\rm d}}(k_x)$ in each layer is given by
\begin{equation}
\begin{gathered}
\tilde{\bm{h}}_1^{{\rm d}}(k_x) = -\frac{1}{\mu_0M_{{\rm s}1}d_1}\bm{g}^\dagger(k_x) \tilde{\bm{m}}_2(k_x), \\
\tilde{\bm{h}}_2^{{\rm d}}(k_x) = -\frac{1}{\mu_0M_{{\rm s}2}d_2}\bm{g}(k_x) \tilde{\bm{m}}_1(k_x),
\end{gathered}
\label{hd}
\end{equation}

\noindent which can be written in the form of
\begin{equation}
\begin{gathered}
\begin{bmatrix}
\tilde{h}^{{\rm d}}_{1x} (k_x) \\
\tilde{h}^{{\rm d}}_{1z} (k_x)
\end{bmatrix} = f_{1}
\begin{bmatrix}
|k_x| {\rm sin}\theta_1 {\rm sin}\theta_2 & i k_x {\rm sin}\theta_1 \\ i k_x {\rm sin}\theta_2 & -|k_x|
\end{bmatrix}
\begin{bmatrix}
\tilde{m}_{2x} (k_x) \\
\tilde{m}_{2z} (k_x)
\end{bmatrix}, \\
\begin{bmatrix}
\tilde{h}^{{\rm d}}_{2x} (k_x) \\
\tilde{h}^{{\rm d}}_{2z} (k_x)
\end{bmatrix} = f_{2}
\begin{bmatrix}
|k_x| {\rm sin}\theta_1 {\rm sin}\theta_2 & -i k_x {\rm sin}\theta_2 \\ -i k_x {\rm sin}\theta_1 & -|k_x|
\end{bmatrix}
\begin{bmatrix}
\tilde{m}_{1x} (k_x) \\
\tilde{m}_{1z} (k_x)
\end{bmatrix},
\end{gathered}
\label{hd components}
\end{equation}

\noindent where $f_{1(2)} = M_{{\rm s}2(1)}d_{2(1)}N(|k_x|;d_1)N(|k_x|;d_2) / 2$.

There are two additional points we would like to mention:

\noindent 1) If the moment distribution along the $z$ direction is a higher-order standing mode within each individual layer, which means $\bm{m}_1 (\bm{r}) =  \bm{m}_1(\bm{r}_\parallel) S_m(z; d_{1})$, $\bm{m}_2 (\bm{r}) =  \bm{m}_2(\bm{r}_\parallel) S_n(z; d_{2})$ with $m, n = 1, 2, ...$, and
\begin{equation}
\begin{gathered}
S_m(z; d_{1}) = \left\{\begin{aligned}
{\rm cos}(m\pi z/d_1) \quad &(-d_1 < z < 0) \\
0 \quad\quad\;\;\quad &({\rm otherwise}),
\end{aligned} \right. \\
S_n(z; d_{2}) = \left\{\begin{aligned}
{\rm cos}(n\pi z/d_2) \quad \; &(0 < z < d_2) \\
0 \quad\quad\;\;\quad \; &({\rm otherwise}),
\end{aligned} \right.
\end{gathered}
\label{higher order S}
\end{equation}

\noindent we can then use the similar analysis as we do for the uniform mode. The only difference is that the factor $N(|k_x|;d_1)$ $N(|k_x|;d_2)$ included in $f_j$ is substituted by $N^m(|k_x|;d_1)$ $N^n(|k_x|;d_2)$, where $N^m(k;d) = kd[1 - (-1)^me^{-kd}]/ [(kd)^2 + (m\pi)^2]$. Therefore, a higher-order standing mode will only decrease the interlayer coupling strength, but will not qualitatively affect the conclusions on the tunable Chern bands and chiral surface spin currents.

\noindent 2) If two layers in Fig. \hyperref[fig:S1]{S1(c)} are separated by a gap of $\delta d$, a factor of ${\rm exp}(-k_\parallel \delta d)$ will be additionally included in $f_j$. Due to this exponential decay character, we only include the dipolar interaction between nearest neighboring layers.

\section{\label{app:B} 2D Magnonic Hamiltonian with only the interlayer dipolar coupling}

We continue focusing on the magnetic bilayers. Since each magnetic moment precesses around its equilibrium orientation, in the small-angle precession limit, we can express the moment as $\bm{m}_{j} (\bm{r}, t) = m_{jx}(\bm{r})e^{i \omega t}\hat{x}_{j} + 1 \cdot \hat{y}_{j} + m_{jz}(\bm{r})e^{i \omega t}\hat{z}$, with the angular frequency $\omega$. The dynamics of precession is characterized by the Landau-Lifshitz (LL) equation
\begin{equation} 
\begin{aligned}
\frac{\partial\bm{m}_{j}}{\partial t} = - \gamma\mu_{0}\bm{m}_{j} \times \bm{h}_{j}
\end{aligned},
\label{LL eq}
\end{equation}

\noindent where $\gamma$ is electron's gyromagnetic ratio, $\mu_{0}$ is vacuum permeability, $\bm{h}_{j} (\bm{r}, t) = h_{jx} (\bm{r}) e^{i \omega t} \hat{x}_{j} + H_{j} \hat{y}_{j} + h_{jz} (\bm{r}) e^{i \omega t} \hat{z}$ is the effective field experienced by each layer, with the static component $H_{j}$ including the external field $H$ and the static interfacial exchange field $H^{\rm ex}_j$. The interfacial exchange energy can be written as \cite{Berkowitz1999}
\begin{equation}
\begin{aligned}
	E^{\rm ex}_{\rm int} &= -J \int \bm{m}_1(\bm{r}_\parallel) \cdot \bm{m}_2(\bm{r}_\parallel) d^2 \bm{r}_\parallel,
\end{aligned}
\label{int ex}
\end{equation}

\noindent where $J$ is the interfacial exchange constant, and the integral is on the whole interface. The static field contributed by the interfacial exchange is therefore derived as $H^{\rm ex}_j = J {\rm cos}(\theta_1 - \theta_2) / (\mu_0 M_{{\rm s}j}d_j)$. In the antiparallel aligned multilayers, i.e., $J < 0$ and $\theta_2 = \theta_1 \pm \pi$, the static interfacial exchange field in each layer is $H^{\rm ex}_j = 2|J| / (\mu_0 M_{{\rm s}j}d_j)$, when including both the interfaces.

Linearizing Eq. \eqref{LL eq} in the small-angle precession limit and transforming it into the reciprocal space, we have
\begin{equation} 
\begin{gathered}
   \frac{i \omega}{\gamma \mu_{0}}\tilde{m}_{jx} = - \tilde{h}_{jz} + H_{j} \tilde{m}_{jz}, \\
   \frac{i \omega}{\gamma \mu_{0}}\tilde{m}_{jz} =  \tilde{h}_{jx} - H_{j} \tilde{m}_{jx}.
\end{gathered}
\label{linear LL eq}
\end{equation}

In the absence of interlayer interactions, each individual layer has its unperturbed eigenfrequencies and eigenstates. The intralayer exchange field is $\bm{h}^{{\rm ex0}}_{j} = [2A^{{\rm ex}}_j / (\mu_{0}M_{{\rm s}j})]\nabla^2\bm{m}_{j}$ \cite{Klingler2014}, with the exchange stiffness constant $A^{{\rm ex}}_j$. In Eq. \eqref{linear LL eq}, this corresponds to:
\begin{equation} 
\tilde{h}^{{\rm ex0}}_{jx(z)} = -\frac{2A^{{\rm ex}}_jk_x^2}{\mu_{0}M_{{\rm s}j}}\tilde{m}_{jx(z)},
\label{intra ex field}
\end{equation}

\noindent The unperturbed eigenfrequencies and eigenstates of each layer are therefore given by
\begin{equation} 
\begin{gathered}
\omega_{j\pm} = \pm(A_jk_x^2 + \Omega_j),\\ 
\tilde{m}_{j\pm} = \frac{1}{\sqrt{2}} (\tilde{m}_{jx} \pm i\tilde{m}_{jz})
,\end{gathered}
\label{eigenfreq}
\end{equation}

\noindent where $A_j = 2A^{{\rm ex}}_{j}\gamma/ M_{{\rm s}j}$ characterizes the intralayer exchange strength and $\Omega_j = \gamma\mu_{0}H_{j}$ is the Larmor precession frequency. From Eq. \eqref{eigenfreq}, we see that the unperturbed eigenstate $\tilde{m}_{j\pm}$ is a right- (left-) handed circularly polarized magnon mode, which corresponds to the positive (negative) eigenfrequency, representing a particle (hole).

Applying the intralayer exchange field [Eq. \eqref{intra ex field}] and interlayer dipolar field [Eq. \eqref{hd components}] into the linearized LL equation [Eq. \eqref{linear LL eq}], we can get the eigenvalue equation
\begin{align} 
\omega
\begin{bmatrix}
\tilde{m}_{1+} \\ \tilde{m}_{2+} \\ \tilde{m}_{1-} \\ \tilde{m}_{2-}
\end{bmatrix}
=
\begin{bmatrix}
\omega_{1+} & \Delta_{1+} & 0 & \delta_{1+} \\
\Delta_{2+} & \omega_{2+} & \delta_{2+} & 0 \\ 
0 & \delta_{1-} & \omega_{1-} & \Delta_{1-} \\ 
\delta_{2-} & 0 & \Delta_{2-} & \omega_{1-}
\end{bmatrix}
\begin{bmatrix}
\tilde{m}_{1+} \\ \tilde{m}_{2+} \\ \tilde{m}_{1-} \\ \tilde{m}_{2-}
\end{bmatrix},
\label{eigen eq kx}
\end{align}

\noindent where the diagonal terms are exactly unperturbed eigenfrequencies [Eq. \eqref{eigenfreq}] and the off-diagonal terms are originated from the interlayer dipolar coupling, which can be written as
\begin{equation}
\begin{gathered}
\Delta_{j\pm} = \frac{1}{2}\gamma\mu_0 f_j[k_x ({\rm sin}\theta_{1}-{\rm sin}\theta_{2}) \pm |k_{x}|(1 - {\rm sin}\theta_{1}{\rm sin}\theta_{2})], \\
\delta_{j\pm} = \frac{1}{2}\gamma\mu_0 f_j[k_x ({\rm sin}\theta_{1}+{\rm sin}\theta_{2}) \mp |k_{x}|(1 + {\rm sin}\theta_{1}{\rm sin}\theta_{2})].
\end{gathered}
\label{A27}
\end{equation}

\noindent Since we are interested in bilayers with antiferromagnetic spin textures with $\theta_{1} =  \pi / 2$, $\theta_{2} =  -\pi / 2$, or $\theta_{1} =  -\pi / 2$, $\theta_{2} =  \pi / 2$, $\delta_{j\pm}$ vanishes and Eq. \eqref{eigen eq kx} becomes block diagonalized.

We can further express the magnonic excitations using the Holstein-Primakoff (HP) transformation \cite{Holstein1940}
\begin{equation} 
\begin{aligned}
\tilde{m}_{j+} \approx \sqrt{\frac{2\gamma\hbar}{M_{{\rm s}j}S_{xy}d_j}}b_{j,k_x}, \quad \tilde{m}_{j-} \approx \sqrt{\frac{2\gamma\hbar}{M_{{\rm s}j}S_{xy}d_j}}b_{j,k_x}^\dagger
\end{aligned}
\label{HP transform}
\end{equation}

\noindent where $S_{xy}$ is the area of the interface, and $b_{j,k_x}$ ($b_{j,k_x}^\dagger$) annihilates (creates) a circularly polarized magnon with wave vector $k_x$ in the $j$th layer. If we define $\bm{b}_{k_x} = [b_{1, k_x}, b_{2, k_x}]^T$, Eq. \eqref{eigen eq kx} can be equivalently written as
\begin{align}
i\frac{d}{dt}
\begin{bmatrix}
\bm{b}_{k_x} \\ \bm{b}_{-k_x}^{\dagger} 
\end{bmatrix}
= \left[\begin{bmatrix} \bm{b}_{k_x} \\ \bm{b}_{-k_x}^{\dagger} \end{bmatrix}, \mathcal{\hat{H}}_{\rm 2L}\right] = \bm{\sigma}_{3}\bm{H}_ {\rm 2L}(k_x)\begin{bmatrix} \bm{b}_{k_x} \\ \bm{b}_{-k_x}^{\dagger} \end{bmatrix},
\label{bilayer H 1}
\end{align}

\noindent with
\begin{equation}
	\begin{gathered} 
	\mathcal{\hat{H}}_{\rm 2L} = \frac{1}{2}\sum_{k_x}\begin{bmatrix} \bm{b}_{k_x}^{\dagger} & \bm{b}_{-k_x} \end{bmatrix} \bm{H}_{\rm 2L}(k_x)\begin{bmatrix} \bm{b}_{k_x} \\ \bm{b}_{-k_x}^{\dagger} \end{bmatrix}, \\
	\bm{\sigma}_{3} \equiv \begin{bmatrix}
	\bm{1} & \bm{0} \\ \bm{0} & -\bm{1} \end{bmatrix}, \quad
	\bm{H}_ {\rm 2L}(k_x) = \begin{bmatrix}
	\bm{h}_ {\rm 2L}(k_x) & \bm{0} \\
	\bm{0} & \bm{h}_ {\rm 2L}^{*}(-k_x) \end{bmatrix},
	\end{gathered}
	\label{bilayer H 2}
\end{equation}

\noindent where
\begin{equation}
\begin{gathered}
\bm{h}_ {\rm 2L}(k_x) = \begin{bmatrix}
\omega_1 & \Delta \\
\Delta & \omega_2 \end{bmatrix}, \\
\omega_j = \omega_{j+} = A_j k_x^2 + \Omega_j, \\
\Delta = \sqrt{\Delta_{1+}\Delta_{2+}} = \frac{1}{2}\gamma\mu_0 \sqrt{f_1 f_2} [k_x ({\rm sin}\theta_{1}-{\rm sin}\theta_{2}) + |k_{x}|(1 - {\rm sin}\theta_{1}{\rm sin}\theta_{2})].
\end{gathered}
\label{bilayer H}
\end{equation}

\noindent The bosonic Bogoliubov-de Gennes (BdG) Hamiltonian $\bm{H}_ {\rm 2L}(k_x)$ only possesses the diagonal blocks, so the bilayer Hamiltonian operator can be simplified as
\begin{equation}
\mathcal{\hat{H}}_{\rm 2L} = \sum_{k_x}\bm{b}_{k_x}^{\dagger} \bm{h}_ {\rm 2L}(k_x)\bm{b}_{k_x}.
\label{bilayer H operator}
\end{equation}

As for the antiparallelly aligned multilayers shown in Fig. 1 of the main text, we can regard it as the stack of bilayers along the $z$ direction. Particularly, when treating the dipolar coupling between two layers within a unit cell, we can set $\theta_1 = \pi/2$, $\theta_2 = -\pi/2$, while we set $\theta_1 = -\pi/2$, $\theta_2 = \pi/2$ for two layers belonging to neighboring cells. Therefore, we can get the intracell (intercell) coupling term $\Delta_{\rm S(D)}$ by assigning the corresponding values of $\theta_{1}$ and $\theta_{2}$ into the expression of $\Delta$ in Eq. \eqref{bilayer H}:
\begin{equation}
\begin{gathered}
\Delta_{\rm S} = \Delta|_{\theta_{1} = \pi / 2, \theta_{2} = -\pi / 2} = B \cdot (|k_{x}| + k_x ), \\
\Delta_{\rm D} = \Delta|_{\theta_{1} = -\pi / 2, \theta_{2} = \pi / 2} = B \cdot (|k_{x}| - k_x ),
\end{gathered}
\label{coupling terms}
\end{equation}

\noindent with $B = \mu_{0}\gamma\sqrt{f_1f_2}$ characterizing the coupling strength. The multilayer magnonic Hamiltonian is therefore written as
\begin{equation}
\begin{aligned}
   \hat{\mathcal{H}} = \hat{\mathcal{H}}_0 + \hat{\mathcal{H}}_{\rm dip} = \sum_{k_x, j, n} \omega_j b_{jn,k_x}^\dagger b_{jn,k_x} + \sum_{k_x, j, n} \left( \Delta_{\rm S} b_{1n,k_x}^\dagger b_{2n,k_x} + \Delta_{\rm D} b_{1n,k_x}^\dagger b_{2,n-1,k_x} + \mathrm{H.c.} \right).
\end{aligned}
\label{multilayer H kx}
\end{equation}
We can get the bulk Hamiltonian [Eq. (1) of the main text], with the periodic boundary condition along the $z$ direction, and we can solve for the explicit surface states [Figs. 2(a) to 2(c) of the main text], with the open boundary condition along the $z$ direction.

\section{\label{app:C} 3D Magnonic Hamiltonian with all interactions included}

In the generic case, magnons can propagate within the whole $xy$ plane in each individual layer, i.e., with the in-plane momenta $\bm{k_\parallel} = k_x \hat{x} + k_y \hat{y}$. Besides, both the intralayer dipolar interaction and the dynamic interfacial exchange interaction should be included.

First, let's derive the interlayer dipolar field with finite $k_y$. For two layers within a unit cell, we have $\theta_1 =  \pi/2$ and $\theta_2 = -\pi/2$. Substituting them into Eq. \eqref{rot matrix} gives
\begin{equation}
\begin{aligned}
R_{1}^\prime = \begin{bmatrix}
1 & 0 & 0 \\
0 & 1 & 0 \\
0 & 0 & 1
\end{bmatrix}, \quad
R_{2}^\prime = \begin{bmatrix}
-1 & 0 & 0 \\
0 & -1 & 0 \\
0 & 0 & 1
\end{bmatrix}.
\end{aligned}
\label{rot matrix prime}
\end{equation}

\noindent Applying these rotation matrices into Eq. \eqref{G kparallel}, we get
\begin{equation}
\begin{aligned}
\bm{g}^\prime(\bm{k}_\parallel) = &(R_2^\prime)^{-1} \bm{G}(\bm{k}_\parallel) R_1^\prime
=\frac{\mu_0M_{{\rm s1}}M_{{\rm s2}}d_1d_2}{2k_\parallel} N(k_\parallel;d_1) N(k_\parallel;d_2)
\begin{bmatrix}
k_x^2 & k_xk_y & -ik_xk_\parallel \\
k_xk_y & k_y^2 & -ik_yk_\parallel \\
ik_xk_\parallel & ik_yk_\parallel & k^2_\parallel
\end{bmatrix}.
\end{aligned}
\label{g prime}
\end{equation}

\noindent Right now, the interlayer dipolar energy is written as
\begin{equation}
\begin{aligned}
E_{{\rm d}}^\prime = &\int \tilde{\bm{m}}_2^{\dagger}(\bm{k}_\parallel) \bm{g}^\prime(\bm{k}_\parallel) \tilde{\bm{m}}_1(\bm{k}_\parallel) \frac{d^2\bm{k}_\parallel}{(2\pi)^2}
= -\mu_0M_{{\rm s}1}d_1 \int \tilde{\bm{h}}_1^{{\rm d}\dagger}(\bm{k}_\parallel)\tilde{\bm{m}}_1(\bm{k}_\parallel) \frac{d^2\bm{k}_\parallel}{(2\pi)^2}
= -\mu_0M_{{\rm s}2}d_2 \int \tilde{\bm{m}}_2^{\dagger}(\bm{k}_\parallel)\tilde{\bm{h}}_2^{{\rm d}}(\bm{k}_\parallel) \frac{d^2\bm{k}_\parallel}{(2\pi)^2},
\end{aligned}
\label{Ed prime}
\end{equation}

\noindent \noindent and the interlayer dipolar field in each layer is give by
\begin{equation}
\begin{gathered}
\tilde{\bm{h}}_1^{{\rm d}}(\bm{k}_\parallel) = -\frac{1}{\mu_0M_{{\rm s}1}d_1}\bm{g}^{\prime\dagger}(\bm{k}_\parallel) \tilde{\bm{m}}_2(\bm{k}_\parallel), \\
\tilde{\bm{h}}_2^{{\rm d}}(\bm{k}_\parallel) = -\frac{1}{\mu_0M_{{\rm s}2}d_2}\bm{g}^\prime(\bm{k}_\parallel) \tilde{\bm{m}}_1(\bm{k}_\parallel),
\end{gathered}
\label{hd kparallel}
\end{equation}

\noindent which can be written in the form of components
\begin{equation}
\begin{aligned}
\begin{bmatrix}
\tilde{h}^{{\rm d}}_{jx} (\bm{k}_\parallel) \\
\tilde{h}^{{\rm d}}_{jz} (\bm{k}_\parallel)
\end{bmatrix} &= f_j^\prime
\begin{bmatrix}
-k_x^2 / k_\parallel & ik_x \\ -i k_x & -k_\parallel
\end{bmatrix}
\begin{bmatrix}
\tilde{m}_{vx} (\bm{k}_\parallel) \\
\tilde{m}_{vz} (\bm{k}_\parallel)
\end{bmatrix}. \\
\end{aligned}
\label{hd kparallel components}
\end{equation}

\noindent where $v = 2$ for $j = 1$ and $v = 1$ for $j = 2$. $f_j^\prime$ can be obtained by replacing $|k_x|$ with $|\bm{k}_\parallel|$ in the expression of $f_j$.

Next, let's derive the intralayer dipolar field, which is similar to what we've done for the interlayer dipolar field. The difference is that there is only one continuous body now, so $\bm{r}$ and $\bm{r^\prime}$ are in the same layer, as shown in Fig. \hyperref[fig:S1]{S1(d)}. The intralayer dipolar energy of the $j$th layer is
\begin{equation}
\begin{aligned}
E_{{\rm d}0,j} = \frac{1}{2}\int \tilde{\bm{m}}_j^{\dagger}(\bm{k}_\parallel) \bm{G}_j(\bm{k}_\parallel) \tilde{\bm{m}}_j(\bm{k}_\parallel) \frac{d^2\bm{k}_\parallel}{(2\pi)^2}.
\end{aligned}
\label{Ed0}
\end{equation}

\noindent $\bm{G}_j({\bm{k}_\parallel})$ is calculated to be
\begin{equation}
\begin{aligned}
\bm{G}_j({\bm{k}_\parallel}) = \frac{1}{3}\mu_0M_{{\rm s}j}^2d_j\bm{1} - \mu_0M_{{\rm s}j}^2d_j\frac{1}{k_\parallel^2}
\begin{bmatrix}
k_x^2(1 - N_j^\prime) & k_xk_y(1 - N_j^\prime) & 0 \\
k_xk_y(1 - N_j^\prime) & k_y^2(1 - N_j^\prime) & 0 \\
0 & 0 & k^2_\parallel N_j^\prime
\end{bmatrix},
\end{aligned}
\label{Gj}
\end{equation}

\noindent where $N_j^\prime  = N(|k_\parallel|;d_j)$. After switching to the local coordinate system, we find $\bm{g}_j(\bm{k}_\parallel) = (R_j^\prime)^{-1} \bm{G}_j(\bm{k}_\parallel) R_j^\prime = \bm{G}_j(\bm{k}_\parallel)$, where $R_j^\prime$ is shown in Eq. \eqref{rot matrix prime}. Right now, the intralayer dipolar energy is written as
\begin{equation}
\begin{aligned}
E_{{\rm d}0, j} = &\frac{1}{2}\int \tilde{\bm{m}}_j^{\dagger}(\bm{k}_\parallel) \bm{g}_j(\bm{k}_\parallel) \tilde{\bm{m}}_j(\bm{k}_\parallel) \frac{d^2\bm{k}_\parallel}{(2\pi)^2}
= -\frac{1}{2}\mu_0M_{{\rm s}j}d_j \int \tilde{\bm{m}}_j^{\dagger}(\bm{k}_\parallel)\tilde{\bm{h}}_j^{{\rm d0}}(\bm{k}_\parallel) \frac{d^2\bm{k}_\parallel}{(2\pi)^2},
\end{aligned}
\label{Ed0j}
\end{equation}

\noindent and the intralayer dipolar field $\tilde{\bm{h}}^{{\rm d0}}_j (\bm{k}_\parallel)$ in each layer is given by
\begin{equation}
\begin{aligned}
\tilde{\bm{h}}_j^{{\rm d0}}(\bm{k}_\parallel) = -\frac{1}{\mu_0M_{{\rm s}j}d_j}\bm{g}_j(\bm{k}_\parallel) \tilde{\bm{m}}_j(\bm{k}_\parallel),
\end{aligned}
\label{hd0j}
\end{equation}

\noindent which can be written in the form of
\begin{equation}
\begin{aligned}
\begin{bmatrix}
\tilde{h}^{{\rm d0}}_{jx} (\bm{k}_\parallel) \\
\tilde{h}^{{\rm d0}}_{jz} (\bm{k}_\parallel)
\end{bmatrix} &= M_{{\rm s}j}
\begin{bmatrix}
(1- N_j^\prime)k_x^2 / k_\parallel^2 + \frac{1}{3}  & 0 \\
0 & N_j^\prime + \frac{1}{3}
\end{bmatrix}
\begin{bmatrix}
\tilde{m}_{jx} (\bm{k}_\parallel) \\
\tilde{m}_{jz} (\bm{k}_\parallel)
\end{bmatrix}.
\end{aligned}
\label{hd0j components}
\end{equation}

On the other hand, from Eq. \eqref{int ex}, we can get the dynamic interfacial exchange field $\tilde{\bm{h}}_j^{{\rm ex}}$ in each layer:
\begin{equation}
\begin{aligned}
\begin{bmatrix}
\tilde{h}^{{\rm ex}}_{jx} (\bm{k}_\parallel) \\
\tilde{h}^{{\rm ex}}_{jz} (\bm{k}_\parallel)
\end{bmatrix} = H^{\rm ex}_j
\begin{bmatrix}
1 / 2 & 0 \\ 0 & - 1 / 2
\end{bmatrix}
\begin{bmatrix}
\tilde{m}_{vx} (\bm{k}_\parallel) \\
\tilde{m}_{vz} (\bm{k}_\parallel)
\end{bmatrix}.
\end{aligned}
\label{B15}
\end{equation}

Applying the intralayer exchange field [replacing $k_x$ by $\bm{k}_\parallel$ in Eq. \eqref{intra ex field}], interlayer dipolar field [Eq. \eqref{hd kparallel components}], intralayer dipolar field [Eq. \eqref{hd0j components}], and dynamic interfacial exchange field [Eq. \eqref{B15}], into the linearized LL equation [Eq. \eqref{linear LL eq}], we can get the eigenvalue equation
\begin{align}
\omega
\begin{bmatrix}
\tilde{m}_{1+} \\ \tilde{m}_{2+} \\ \tilde{m}_{1-} \\ \tilde{m}_{2-}
\end{bmatrix}
=
\begin{bmatrix}
\omega_{1}^\prime & \Delta_{1+}^\prime & \delta_1 & \delta_{1}^\prime \\
\Delta_{2+}^\prime & \omega_{2}^\prime & \delta_{2}^\prime & \delta_2 \\ 
-\delta_1 & -\delta_{1}^\prime & -\omega_{1}^\prime & \Delta_{1-}^\prime \\ 
-\delta_{2}^\prime & -\delta_2 & \Delta_{2-}^\prime & -\omega_{2}^\prime
\end{bmatrix}
\begin{bmatrix}
\tilde{m}_{1+} \\ \tilde{m}_{2+} \\ \tilde{m}_{1-} \\ \tilde{m}_{2-}
\end{bmatrix},
\label{eigen eq kparallel}
\end{align}

\noindent with
\begin{equation}
\begin{gathered}
   \omega_{j}^\prime = A_j\bm{k}_\parallel^2 + \Omega_j + \frac{\gamma\mu_0M_{{\rm s}j}}{2}\left[(1 - N_{j}^\prime)\frac{k_x^2}{k_\parallel^2} + N_{j}^\prime + \frac{2}{3}\right], \\
   \Delta_{j\pm}^\prime = \frac{1}{2}\gamma\mu_0 f_j^\prime\left[\pm \left(k_\parallel + \frac{k_x^2}{k_\parallel}\right) + 2k_x\right], \\
   \delta_{j} = \frac{\gamma\mu_0M_{{\rm s}j}}{2} \left[(1 - N_{j}^\prime)\frac{k_x^2}{k_\parallel^2} - N_{j}^\prime\right], \\
   \delta_{j}^\prime = \frac{\gamma\mu_0}{2} \left[f_j^\prime \cdot \left( k_\parallel - \frac{k_x^2}{k_\parallel} \right) + H^{\rm ex}_j\right].
\end{gathered}
\label{elements in eigen eq}
\end{equation}

\noindent Eqs. \eqref{eigen eq kparallel} and \eqref{elements in eigen eq} give the eigenvalue equation for two layers within the same unit cell ($\theta_1 =  \pi/2$ and $\theta_2 = -\pi/2$). As for two layers belonging to neighboring cells ($\theta_1 =  -\pi/2$ and $\theta_2 = \pi/2$), the only difference in the eigenvalue equation is
\begin{equation}
    \Delta_{j\pm}^\prime = \frac{1}{2}\gamma\mu_0 f_j^\prime\left[\pm \left(k_\parallel + \frac{k_x^2}{k_\parallel}\right) - 2k_x\right].
\label{different coupling terms}
\end{equation}

Then we can do the HP transformation on $\tilde{m}_{j+}$ and $\tilde{m}_{j-}$ [Eq. \eqref{HP transform}] and apply the periodic boundary condition along the $z$ direction. We finally get the bulk magnonic Hamiltonian for the 3D multilayers [Eqs. (3) to (6) of the main text].

\bibliographystyle{apsrev4-2}
\bibliography{sm}